\documentclass[12pt]{article}
\usepackage{graphics}

\newcommand\lapproxeq{\lower .7ex\hbox{$\;\stackrel{\textstyle<}{\sim}\;$}}
\newcommand\gapproxeq{\lower .7ex\hbox{$\;\stackrel{\textstyle>}{\sim}\;$}}
\newcommand\alps{\alpha_s}
\newcommand\kt{k_{\perp}}
\newcommand\ycut{y_{\mathrm{cut}}}
\newcommand\GF{\Phi_{\mathrm{jet}}}
\newcommand\G{\Phi}
\newcommand\etam{\eta_{\mathrm{max}}}
\newcommand\cO{\mathcal{O}}
\newcommand\cM{\mathcal{M}}

\begin{document}
\titlepage
\begin{flushright}
CERN--TH/99--239 \\
RAL--TR--1999--049\\
MC--TH--99/10 \\
August 1999 \\
\end{flushright}

\begin{center}
\vspace*{2cm}
{\Large \bf
Subjet Rates in Hadron Collider Jets%
} \\
\vspace*{1cm}

J.R.~Forshaw$^1$\footnote{On leave of absence from the Department of Physics
and Astronomy, University of Manchester, Manchester.  M13 9PL\@.  England.}
and M.H.~Seymour$^2$

\vspace*{0.5cm}

$^1$ Theory Division, CERN, \\ 1211 Geneva 23, Switzerland. \\
$^2$ Particle Physics Department,\\ Rutherford Appleton Laboratory, \\
Chilton, Didcot, Oxon.  OX11 0QX\@.  England.
\end{center}

\vspace*{3cm}
\begin{abstract}
We calculate the subjet rates for jets produced in hadron collisions.
The $\kt$ algorithm is used to define the jets and allows the theoretical
calculation to sum both the leading and next-to-leading
logarithms in the resolution variable, $\ycut$.  We also ensure that our
calculation matches exactly the leading order in $\alps$ result and has
sensible behaviour near thresholds.
\end{abstract}

\newpage

\section{Introduction}
Hadron collisions are a prolific source of hadronic jets.  When these
are hard and well-separated they can be used for a variety of
high-precision tests of perturbative QCD~\cite{hardQCD}.  However, much
of the theoretical interest comes from the soft region in which the
confinement of the produced quarks and gluons into the observed hadrons
plays a crucial r\^ole.  Since this is outside perturbative control a
number of models have been proposed to describe it.  Although these all
give reasonable descriptions of hadron-level data, to gain a deeper
understanding one would like to bridge the gulf between the individual
hadrons and the hard well-separated jets, in such a way that one can
choose the dominant domain, smoothly moving between the two extremes.  A
powerful way of doing this is to define {\em subjets\/} within the
jets~\cite{subjet}.

Subjets are defined by taking all the particles that ended up in a given
jet and running a jet algorithm on those particles with a dimensionless
resolution scale, $\ycut$.  For large $\ycut$, every jet consists of
only a single subjet.  As $\ycut$ is decreased the fraction of jets that
are resolved into two or more subjets increases, until for small $\ycut$
every jet consists of many subjets.  Eventually for small enough
$\ycut$, every final-state particle is resolved from every other, and
subjet distributions become identical to hadron distributions.  The most
interesting region is the intermediate one in which the resolution scale
is large enough for perturbation theory to be valid, but small enough
for the typical multiplicity of subjets to be large.  Careful study of
this region already shows many of the features of the hadronic final
state, despite being fully calculable in perturbation theory.

Subjet studies are already common in $e^+e^-$ annihilation\footnote{Note
that the nomenclature we use is that of hadron collisions in which one
imagines a two-step procedure: first defining jets, then studying
subjets within them.  In $e^+e^-$ annihilation, this nomenclature is
sometimes used, for example in studying the subjet structure of
three-jet events, but since $e^+e^-$ annihilation is a {\em
point-like\/} source of two-jet events, we can consider every jet study
there to be a subjet study, since the first step is not
needed.}~\cite{subjet,e+e-,DO}.  However in hadron collisions there have
been relatively few studies, mainly because most experiments use
cone-based jet algorithms, which are not amenable to the all-orders
calculations that are needed to make subjets interesting.  However, with
the advent of the $\kt$ algorithm for hadron collisions~\cite{ES,CDSW},
subjet studies became possible~\cite{MHS}.  $\kt$ algorithms are of the
clustering type and are defined by a {\em closeness measure}, which
decides whether two particles are clustered together, and a {\em
recombination scheme}, which specifies how they are clustered together.
Once these two details have been defined, the algorithm can be applied
iteratively to final states containing any number or type
(partons/hadrons/calorimeter cells) of particle.  For collisions
involving incoming hadron beams, one also has to define a closeness to
the beam direction to ensure that the resulting cross sections obey the
factorization theorem~\cite{CDW}.  The variant we use is the `inclusive
$\kt$ algorithm' with the `$p_t$-recombination scheme'.

To be precise, we define jets in the following way, in terms of a
dimensionless parameter $R$, which we will see plays a radius-like
r\^ole in defining the angular extent of the jets, and which we assume
is of order~1.  We start with a list of `particle' momenta that could
consist of partons, hadrons or calorimeter cells, and an empty list of
jet momenta.  In the $p_t$ scheme, all particles are treated as if they
were massless, so momenta are defined by only three variables,
transverse momentum\footnote{Note that the transverse momentum, $p_t$,
is often replaced by the transverse energy,~$E_T$.  Since the
$p_t$-scheme uses massless kinematics, the two are identical.},~$p_t$,
pseudorapidity,~$\eta$, and azimuth,~$\phi$.  Since these are invariant
under Lorentz boosts along the beam direction, the whole algorithm is
longitudinally boost invariant.%
\vspace{-\parskip}
\begin{enumerate}
  \addtolength{\itemsep}{-\parskip}
  \item For every pair of particles, $i,j$, calculate their closeness
  \begin{equation}
    d_{ij} = \min(p_{t,i}^2,p_{t,j}^2)\;\Delta R_{ij}^2,
    \quad\left(\Delta R_{ij}^2=(\eta_i-\eta_j)^2+(\phi_i-\phi_j)^2\right).
  \end{equation}
  Note that in the small angle limit, this reduces to the transverse
  momentum of the softer particle relative to the axis defined by the
  harder:
  \begin{equation}
    d_{ij} \approx {\kt}_{ij}^2
    \equiv \min(E_i^2,E_j^2)\;\sin^2\Delta\theta_{ij},
    \quad \Delta R_{ij}^2\ll1.
  \end{equation}
  \item For every particle, $i$, calculate its closeness to the beam
  directions
  \begin{equation}
    d_i = p_{t,i}^2\;R^2.
  \end{equation}
  \item Find the smallest closeness
  $d_{\mathrm{min}}=\min\left\{d_{ij},d_i\right\}$.
  \begin{enumerate}
    \item If $d_{\mathrm{min}}=d_{ij}$, particles $i$ and $j$ are
    merged, by replacing their entries in the particle list by a new
    entry, with momentum
    \begin{eqnarray}
      p_{t,ij}  &=& p_{t,i} + p_{t,j}, \\
      \eta_{ij} &=& \Bigl(p_{t,i}\eta_i  + p_{t,j}\eta_j\Bigr)/p_{t,ij},  \\
      \phi_{ij} &=& \Bigl(p_{t,i}\phi_i  + p_{t,j}\phi_j\Bigr)/p_{t,ij}.
    \end{eqnarray}
    \item If $d_{\mathrm{min}}=d_i$, $i$ is said to be a completed
    jet: its momentum is moved from the list of particles to the list
    of jets.
  \end{enumerate}
  \item Continue from step~1 until the list of particles is empty.
\end{enumerate}
\vspace{-2\parskip}
Note that there is an unambiguous assignment of every final-state
particle to exactly one completed jet.  Note also that although the
particles are merged in order of increasing relative transverse
momentum, it is actually their angular separation that controls whether
or not they are merged: every merged pair within a jet has $\Delta
R_{ij}<R$ and every jet is separated from every other by $\Delta
R_{ij}>R$.

Having selected a particular jet $k$ for further study, we define
subjets within it, in terms of a dimensionless parameter $\ycut$, by
rerunning the above algorithm, but only putting those particles that
were assigned to our jet in the initial list, and stopping as soon
as\footnote{Note that it is possible to reach a situation in which,
although (\ref{ycutdef}) is satisfied, further merging would result in
a configuration in which it was no longer satisfied, and only even later
in the merging history would it again be satisfied.  Such
`non-monoticity' is a property of the recombination scheme, and is
shared by the only other scheme in common use, the $E$-scheme.  It is
possible to define more complicated schemes that are guaranteed to be
monotonic, like the $p_t^2$-scheme, and the results in \cite{CDSW}
show that, as one might expect, they suffer smaller hadronization
corrections.} all closenesses satisfy
\begin{equation}
  \label{ycutdef}
  d_{ij} > \ycut \; p_{t,k}^2.
\end{equation}
All particles in the list are then called subjets.

Experimental implementations of this algorithm (see for example
\cite{DZ}) have tended to also define a preclustering stage in
which the calorimeter cells are clustered together with a fixed angular
cut.  We do not consider the effect of such a procedure here, so cannot
directly compare with their data at present.
In \cite{DZ}, $d_{ij}$ and $d_i$ are also defined slightly differently:
both are divided by a factor of $R^2$ relative to ours.  $\ycut$ is
still defined through (\ref{ycutdef}) though, so that it is different
from ours by a factor of $R^2$.  This does not affect any physical
results, except that their $\ycut$ is therefore identical to our $Y$,
defined in (\ref{Ydef}).

In \cite{MHS} the mean number of subjets in a hadron collider jet was
calculated, to leading order in $\alps$ and summing leading and
next-to-leading logarithms of $\ycut$ to all orders in $\alps$.  In this
paper we go one step further by calculating the $n$-subjet rates,
$P_n(\ycut)$, i.e.~the fraction of jets that consist of $n$ subjets at a
given resolution scale, to the same accuracy.

In general, one finds that $P_n\sim\alps^{n-1}$ and that at small
$\ycut$ the leading term at each order $\alps^m$, $m\ge n-1$, is
$\ln^{2m}\ycut$.  In the region we are interested in
$\alps\ln^2\ycut\gapproxeq1$ and order-by-order perturbation theory does
not converge, so these (leading-logarithmic) terms must be summed to all
orders in $\alps$.  Since each higher order in $\alps$ contributes two
additional logarithms, we must also sum the next-to-leading logarithms,
$\alps^m\ln^{2m-1}\ycut$, before we obtain a result that is of
leading-order accuracy (i.e.~in which neglected terms are down by a
power of~$\alps$, without any logarithmic enhancement, relative to
included terms).  The modified leading logarithmic approximation (MLLA)
provides a framework to perform this next-to-leading logarithmic
resummation (see for example \cite{DKMT,ESW}).  It turns out that
although the leading logarithms are identical in hadron collisions to
the well-studied case of $e^+e^-$ annihilation, the next-to-leading
logarithms contain essential contributions from soft gluons that are
radiated off the incoming partons.  The probability of this depends on
the jet kinematics, the parton distribution functions, the collision
type and energy and so forth, so cannot easily be treated analytically.
However, as pointed out in \cite{MHS}, it is possible to manipulate
all these effects into a single factor, which can be calculated
numerically, multiplying a tower of logarithms that can be summed
analytically.

Finally, in order to describe the region of large $\ycut$, it is
necessary to include in $P_n$ the exact tree-level contributions from
the $n+1$-parton final state.  Although these are available for $n$ up
to~6~\cite{NJETS}, we have only included them up to $n=2$.  Formally,
our accuracy is therefore
\begin{equation}
  \label{accuracy}
  \begin{array}{rclr}
  (1-P_1) &=& (1-P_1)\left(1+\cO(\alps)\right) & \quad\forall\;\ycut, \\
     P_2  &=&    P_2 \left(1+\cO(\alps)\right) & \quad\forall\;\ycut, \\
  P_{n>2} &=& P_{n>2}\left(1+\cO(\alps)\right) & \quad\ycut\ll1.
  \end{array}
\end{equation}
Since the higher subjet rates are small for large $\ycut$, this is not a
severe limitation.

Within the MLLA the flavour of a jet is well-defined, because all
emission is either of a soft gluon, or a collinear parton pair, both of
which preserve the jet's net flavour.  However, after including the
exact $\cO(\alps)$ contribution this is no longer the case as, for
example, configurations arise in which the two subjets of a jet are both
quarks.  As shown in \cite{MHS} this is a tiny effect
and for all practical purposes quark and gluon jets can be separated.
Thus for all our results we show three distributions: for the sum of all
jet types (which include the unclassifiable jets), and for quark and
gluon jets separately.  In fact one of our main results is the fact that
the properties of quark and gluon jets of a given transverse momentum
are almost entirely insensitive to how they were created: their position
in the detector, the beam particles and energies, etc\footnote{This is
no longer true, however, if one breaks the inclusive nature of the jet
definition, by imposing cuts on the other jets in the event.  Thus
although we would expect our results to describe inclusive
photoproduction data such as \cite{ZEUS} reasonably well, we would
expect them to break down
once an $x_\gamma$ cut is imposed, as it often is in experimental
analyses.  We consider this, and other contributions unique to
photoproduction, in a future publication.}.

\section{Summing the leading and next-to-leading logarithms}
As mentioned in the introduction, the leading logarithms are identical
in hadron collisions to in $e^+e^-$ annihilation, but the
next-to-leading logarithms receive additional contributions from
initial-state radiation.  We therefore first consider the final-state
contributions in Section~\ref{finallogs} before adding in the
initial-state contributions in Section~\ref{inilogs}.  This in turn leads
us to a natural way of matching with the exact $\cO(\alps)$
matrix-element result, which we discuss in Section~\ref{fixed}.
All of these results can be improved by an improved treatment of the
threshold region in the resummed calculation, which we discuss in
Section~\ref{thresh}.

\subsection{Final-state logarithms}
\label{finallogs}
Within the MLLA, the hardness of a jet is characterised by the product
of its energy and the maximum allowed angle of emission from it, in our
case $Q=p_t\,R$.  Its subjet structure is resolved with respect to a
cutoff on transverse momentum $Q_0=p_t\,\surd\ycut$.  We seek to sum the
leading and next-to-leading logarithms of $Q/Q_0$ to all orders.  To
this accuracy, we are insensitive to the precise values of $Q$ and
$Q_0$ provided $Q/Q_0$ is left unchanged, i.e.~we are free to multiply
both by any number of order unity.  It will turn out to be useful to
have $Q=p_t$, and hence we define
\begin{eqnarray}
  Q &=& p_t, \\
  \label{Ydef}
  Q_0 &=& p_t\,\surd\ycut/R \equiv p_t \,\surd Y.
\end{eqnarray}
Our procedure correctly sums logarithms of $Y$ to all orders in $\alps$,
but does not keep track of the additional logarithms that would arise if
$R$ was much less than~1.

Considering only final-state logarithms, the probability of finding $n$
subjets within a jet of flavour $a$, $P_n^a(Q,Y)$, does not depend on
how the jet was produced.  In order to compute it we shall use a
generating function, $\G_a(u,Q)$, defined by
\begin{equation}
\G_a(u,Q) = \sum_{n=1}^{\infty} u^n P_n^a(Q,Y),
\end{equation}
from which it follows that
\begin{equation}
P_n^a(Q,Y) = \left. \frac{1}{n!} \frac{d^n \G_a(u,Q)}{du^n} \right|_{u=0}.
\end{equation}
For brevity, we suppress the dependence of the generating function on
$Y$.  The use of generating functions in describing quark and gluon
jets in $e^+e^-$ collisions is now textbook material, see for example
\cite{DKMT,ESW}.  They obey the coupled evolution
equations~\cite{DO,DKMT,ESW}:
\begin{eqnarray}
  \label{master}
  \G_a(u,Q) &=& u +
  \frac12\sum_b\int_{4Q_0^2}^{Q^2} \frac{dQ'^2}{Q'^2}
  \int_{Q_0/Q'}^{1-Q_0/Q'} dz
  \nonumber\\&&\qquad
  \frac{\alps(z(1-z)Q')}{2\pi} P_{a\to bc}(z)
  \biggl(\G_b(u,zQ')\,\G_c(u,(1-z)Q') - \G_a(u,Q')\biggr),
  \phantom{(99)}
\end{eqnarray}
where $P_{a\to bc}(z)$ is the leading order DGLAP splitting kernel for
the branching $a\to b+c$, with $b$ taking an energy fraction $z$.  In
the final bracket of (\ref{master}), the first term comes from the
real splitting $a\to b+c$, while the second term comes from the
corresponding virtual terms, such that unitarity is preserved in the
sum of the two.

These equations can be rewritten in a form in which they are easier to
solve, by introducing the Sudakov form factors,
\begin{equation}
  \label{sudakov}
  \Delta_a(Q) = \exp\left\{-
  \frac12\sum_b\int_{4Q_0^2}^{Q^2} \frac{dQ'^2}{Q'^2}
  \int_{Q_0/Q'}^{1-Q_0/Q'} dz
  \frac{\alps(z(1-z)Q')}{2\pi} P_{a\to bc}(z)
  \right\},
\end{equation}
which sum the virtual terms to all orders, to give
\begin{eqnarray}
  \label{master2}
  \G_a(u,Q) &=& u\Delta_a(Q)
  \exp\Biggl\{
  \frac12\sum_b\int_{4Q_0^2}^{Q^2} \frac{dQ'^2}{Q'^2}
  \int_{Q_0/Q'}^{1-Q_0/Q'} dz
  \nonumber\\&&\qquad
  \frac{\alps(z(1-z)Q')}{2\pi} P_{a\to bc}(z)
  \frac{\G_b(u,zQ')\,\G_c(u,(1-z)Q')}{\G_a(u,Q')}
  \Biggr\}.
  \phantom{(99)}
\end{eqnarray}
Note that the integration has soft singularities at $z=0$ and $z=1$.  It
is easier to solve (\ref{master2}) if we rewrite it in such a way
that all the singularities appear at $z=0$.
We multiply the integrand by $1=z+(1\!-\!z)$ and use the fact that
$P_{a\to bc}(z) = P_{a\to cb}(1-z)$ to obtain
\begin{eqnarray}
  \label{master2c}
  \G_a(u,Q) &=& u\Delta_a(Q)
  \exp\Biggl\{
  \sum_b\int_{4Q_0^2}^{Q^2} \frac{dQ'^2}{Q'^2}
  \int_{Q_0/Q'}^{1-Q_0/Q'} dz
  \nonumber\\&&\qquad
  \frac{\alps(z(1-z)Q')}{2\pi} \;(1-z)P_{a\to bc}(z)
  \frac{\G_b(u,zQ')\,\G_c(u,(1-z)Q')}{\G_a(u,Q')}
  \Biggr\}.
  \phantom{(99)}
\end{eqnarray}
This equation only has a soft singularity at $z=0$.  We are therefore
free to replace $(1\!-\!z)$ by 1 in all smoothly-varying functions, and
to rewrite the upper limit of the $z$ integration as~1.
The form of the soft singularity is universal so, introducing
$C_{a=q}=C_F$ and $C_{a=g}=C_A$, we can extract it:
\begin{eqnarray}
  \label{master2d}
  \G_a(u,Q) &=& u\Delta_a(Q)
  \exp\Biggl\{
  \int_{Q_0^2}^{Q^2} \frac{dQ'^2}{Q'^2}
  \int_{Q_0/Q'}^1 dz \;\frac{\alps(zQ')}{2\pi}
  \nonumber\\&&
  \Biggl(\frac{2C_a}z\G_g(u,zQ')+
  \sum_b\left[(1-z)P_{a\to bc}(z)
  -\delta_{bg}\frac{2C_a}z\right]
  \frac{\G_b(u,zQ')\,\G_c(u,Q')}{\G_a(u,Q')}\Biggr)
  \Biggr\}.
  \phantom{(99)}
\end{eqnarray}
Since we have removed the soft singularity, we can replace $z$ by 1 in
the final term.  Rewriting somewhat, we therefore obtain:
\begin{eqnarray}
  \label{master4}
  \G_a(u,Q) &=& u\Delta_a(Q)
  \exp\Biggl\{
  \int_{Q_0^2}^{Q^2} \frac{dQ'^2}{Q'^2}
  \frac{\alps(Q')}{2\pi} \Biggl(
  C_a
  \ln\frac{Q^2}{Q'^2}\,
  \G_g(u,Q')
  \nonumber\\&&\qquad
  +
  \sum_b
  \frac{\G_b(u,Q')\,\G_c(u,Q')}{\G_a(u,Q')}
  \int_{Q_0/Q'}^1 dz
  \left((1-z)P_{a\to bc}(z)-\delta_{bg}\frac{2C_a}{z}\right)
  \Biggr)
  \Biggr\}.
  \phantom{(99)}
\end{eqnarray}

Finally, the lower limit of the final $z$ integration can be replaced
by~0.  Writing these equations explicitly, we therefore have
\begin{eqnarray}
\label{masterq}
\G_q(u,Q) &=& u \Delta_q(Q) \exp \left[ \int_{Q_0^2}^{Q^2}  dQ'^2 \
\Gamma_q(Q,Q') \G_g(u,Q') \right], \\
\label{masterg}
\G_g(u,Q) &=& u \Delta_g(Q) \exp \Bigg[ \int_{Q_0^2}^{Q^2}  dQ'^2 \
\Big\{ \Gamma_g(Q,Q') \G_g(u,Q') + \Gamma_f(Q')
\frac{\G_q(u,Q')^2}{\G_g(u,Q')} \Big\} \Bigg],
\end{eqnarray}
and
\begin{eqnarray}
\Delta_g(Q) &=& \exp \left[ -\int_{Q_0^2}^{Q^2} dQ'^2
( \Gamma_g(Q,Q')+\Gamma_f(Q')) \right], \label{e9} \\
\Delta_q(Q) &=& \exp \left[ -\int_{Q_0^2}^{Q^2} dQ'^2 \Gamma_q(Q,Q') \right],
\label{e10}
\end{eqnarray}
where we have defined
\begin{eqnarray}
\Gamma_f(Q) &=& \frac{N_f}{6 \pi} \frac{\alps(Q^2)}{Q^2}\,, \label{e11} \\
\Gamma_g(Q,Q') &=& \frac{C_A}{2 \pi} \frac{\alps(Q'^2)}{Q'^2} \left( \ln
\left(\frac{Q^2}{Q'^2}\right)- \frac{11}{6} \right), \label{e12} \\
\Gamma_q(Q,Q') &=& \frac{C_F}{2 \pi} \frac{\alps(Q'^2)}{Q'^2} \left( \ln
\left(\frac{Q^2}{Q'^2}\right)- \frac{3}{2} \right). \label{e13}
\end{eqnarray}
Throughout this paper, we use the one-loop renormalization group
equation for the running of $\alps$ within a given jet, but with the
starting value fixed by the input parton distribution function set
(two-loop) at scale $Q=p_t$:
\begin{equation}
\alps(Q^2) = \frac{\alps^{\mathrm{nlo}}(p_t^2)}
{1 + \frac{\alps^{\mathrm{nlo}}(p_t^2)}{4\pi}
\,\beta_0\ln\frac{Q^2}{p_t^2}}\,,
\end{equation}
where $\beta_0 \equiv (11 C_A - 2 N_f)/3$.

The most convenient way to solve (\ref{masterq},\ref{masterg}) is
by power series expansion in $u$.  Writing
\begin{equation}
\G_a(u,Q) = \sum_{n=1}^{\infty} u^n \G_a^{(n)}(Q), \label{e1}
\end{equation}
we find
\begin{eqnarray}
\G_q^{(1)}(Q) &=& \Delta_q(Q),  \label{e2} \\
\G_g^{(1)}(Q) &=& \Delta_g(Q),  \label{e3}
\end{eqnarray}
\begin{eqnarray}
\G_q^{(2)}(Q) &=& \Delta_q(Q) \int_{Q_0^2}^{Q^2} dQ'^2 \
\Gamma_q(Q,Q') \Delta_g(Q'), \label{e4}  \\
\G_g^{(2)}(Q) &=& \Delta_g(Q) \int_{Q_0^2}^{Q^2} dQ'^2
\left[ \Gamma_g(Q,Q') \Delta_g(Q') + \Gamma_f(Q')
  \frac{\Delta_q(Q')^2}{\Delta_g(Q')} \right], \label{e5}
\end{eqnarray}
\begin{eqnarray}
\G_q^{(3)}(Q) &=& \frac{{\G_q^{(2)}(Q)}^2}{2 \Delta_q(Q)} + \Delta_q(Q)
\int_{Q_0^2}^{Q^2} dQ'^2 \ \Gamma_q(Q,Q') \G_g^{(2)}(Q'), \label{e6} \\
\G_g^{(3)}(Q) &=& \frac{{\G_g^{(2)}(Q)}^2}{2 \Delta_g(Q)} + \Delta_g(Q)
\int_{Q_0^2}^{Q^2} dQ'^2 \ \Bigg\{ \Gamma_g(Q,Q') \G_g^{(2)}(Q') \nonumber
\\ & & + \left. \Gamma_f(Q') \frac{\Delta_q(Q')}{\Delta_g(Q')} \left[
2 \G_q^{(2)}(Q') - \frac{\Delta_q(Q')}{\Delta_g(Q')} \G_g^{(2)}(Q') \right]
\right\}, \label{e7}
\end{eqnarray}
\begin{eqnarray}
\G_q^{(4)}(Q) &=& -\frac{{\G_q^{(2)}(Q)}^3}{3 \Delta_q(Q)^2} + \Delta_q(Q)
\int_{Q_0^2}^{Q^2} dQ'^2 \ \Gamma_q(Q,Q') \G_g^{(3)}(Q')
+ \frac{\G_q^{(2)}(Q) \G_q^{(3)}(Q)}{\Delta_q(Q)}\,, \label{e7p} \\
\G_g^{(4)}(Q) &=& -\frac{{\G_g^{(2)}(Q)}^3}{3 \Delta_g(Q)^2} + \Delta_g(Q)
\int_{Q_0^2}^{Q^2} dQ'^2 \ \Bigg[ \Gamma_g(Q,Q') \G_g^{(3)}(Q') +
\frac{\Gamma_f(Q')}{\Delta_g(Q')} \Bigg( \ \G_q^{(2)}(Q')^2
\nonumber \\ & & \hspace*{-2cm}
+ \ 2 \Delta_q(Q') \G_q^{(3)}(Q') -
\frac{\Delta_q(Q')^2}{\Delta_g(Q')} \G_g^{(3)}(Q') - 2
\frac{\Delta_q(Q')}{\Delta_g(Q')} \G_q^{(2)}(Q') \G_g^{(2)}(Q') +
\frac{\Delta_q(Q')^2}{\Delta_g(Q')^2} \G_g^{(2)}(Q')^2 \Bigg)  \Bigg]
\nonumber \\ &+&
\frac{\G_g^{(2)}(Q) \G_g^{(3)}(Q)}{\Delta_g(Q)}\,.
\label{e8}
\end{eqnarray}
These results sum the large leading ($\alps^m\ln^{2m}Y$) and
next-to-leading ($\alps^m\ln^{2m-1}Y$) logarithms to all orders in
$\alps$.  However they are not guaranteed to be reliable, or even
physically-behaved, in the threshold region $Y\sim1$ where logarithms of
$Y$ do not dominate.  We discuss an improved treatment of this region in
the next section.

Having these generating functions at our disposal it is straightforward
to calculate the $n$-subjet rates in hadron collisions in the
`final-state approximation' (we discuss corrections arising from initial
state radiation in Section~\ref{inilogs}). The jet generating function $\GF$
is formed from the following admixture of the two parton types:
\begin{equation}
\label{admixture}
\GF(u,Q) = \sum_a F_a \G_a(u,Q),
\end{equation}
where the factor $F_a$ labels the fraction of events that lead to the
production of a leading parton of type $a$ ($a=q$ or $g$), i.e.
\begin{equation}
F_a = \frac{\int_{-\etam}^{\etam} d \eta' \frac{1}{s \hat{s}} f_1(x_1) f_2(x_2)
|\cM_2^a|^2 }{\int_{-\etam}^{\etam} d \eta' \frac{1}{s \hat{s}}
f_1(x_1) f_2(x_2) |\cM_2|^2 }.
\end{equation}
Here $f_i(x_i)$ denotes the parton distribution function of hadron $i$
(the factorization scale dependence and flavour dependence are both
suppressed), $\eta'$ is the rapidity of the recoiling parton; it is
integrated over all phase space, i.e.~$\etam = \ln (\sqrt{s}/p_t)$.
The Mandelstam $s$ is that of the hadron-hadron system and $\hat{s}$ is that of
the parton-initiated subprocess, i.e.~$\hat s = x_1 x_2 s$.
The $2 \to 2$ matrix elements are computed to lowest order in $\alps$.

\subsection{Threshold region}
\label{thresh}
The solutions to (\ref{masterq},\ref{masterg}),
i.e.~equations~(\ref{e2}--\ref{e8}), obey the physical boundary
condition that at
$Y=1$ every jet consists of 1~subjet.  However, reducing $Y$, one finds
that $\G_a^{(1)}$ becomes greater than one, while $\G_a^{(2)}$ becomes
negative, which is clearly unphysical.  Only at considerably smaller $Y$
do they start to become physically behaved.  Formally this is not a
problem, as our results are only strictly valid for very small $Y$.
To get a good description of the large $Y$ region one needs to
match with the exact order-by-order perturbative results.

Phenomenologically however, it is highly desirable for the resummed
results to be physically behaved for all $Y$ and to have thresholds at
the correct point in $Y$.  That way we expect the matching with
fixed-order perturbation theory to be smoother, and to obtain reliable
results at a lower perturbative order.

There are two issues to address: the position of the threshold and the
behaviour close to it.  We tackle them in reverse order.

On critical examination of the sequence of approximations we made to
solve (\ref{master2}), we find that all obey positive-definiteness
of the probability distributions except the very last: replacing the
lower limit of the $z$ integration of the non-soft splitting functions
by zero.  Up to and including (\ref{master4}), the soft and non-soft
parts of the splitting functions are treated on an equal footing, but in
getting from (\ref{master4}) to (\ref{masterq},\ref{masterg}),
we apply looser phase space limits to the non-soft terms than to the
soft terms.  Since the non-soft splitting functions are not positive
definite, it is not surprising that extending the range over which they
are integrated, without simultaneously increasing the range over which
the soft ones are integrated, leads to negative distributions.

A straightforward solution to this unphysical behaviour was
proposed\footnote{Our solution is slightly different from theirs: where
  we have $\int_{Q_0/Q'}^1 dz \, P^{\mathrm{non-soft}}_{a\to bc}(z)$
  they have $\int_{Q_0/2Q'}^{1-Q_0/2Q'} dz \,
  P^{\mathrm{non-soft}}_{a\to bc}(z)$.  Although their form also gives a
  physically-behaved threshold at $Y=1$, it still does not treat the
  soft and non-soft terms equivalently.  As a result it is not as good
  an approximation of the full result.} in
\cite{DO}: one solves (\ref{master4}) without approximating the
$z$ limit.  That is, one simply replaces the $\Gamma$ functions by
\begin{eqnarray}
\label{Gfmod}
\Gamma_f(Q;Q_0) &=& \frac{N_f}{2\pi} \frac{\alps(Q^2)}{Q^2}
\left[\frac13
-{\textstyle{\left(\frac{Q_0}{Q}\right)}}
+\frac32{\textstyle{\left(\frac{Q_0}{Q}\right)^2}}
-\frac43{\textstyle{\left(\frac{Q_0}{Q}\right)^3}}
+\frac12{\textstyle{\left(\frac{Q_0}{Q}\right)^4}}
\right], \\
\hspace*{-1em}
\label{Ggmod}
\Gamma_g(Q,Q';Q_0) &=& \frac{C_A}{2\pi} \frac{\alps(Q'^2)}{Q'^2} \left( \ln
\left(\frac{Q^2}{Q'^2}\right)-
\left[\frac{11}{6}
-4{\textstyle{\left(\frac{Q_0}{Q'}\right)}}
+3{\textstyle{\left(\frac{Q_0}{Q'}\right)^2}}
-\frac43{\textstyle{\left(\frac{Q_0}{Q'}\right)^3}}
+\frac12{\textstyle{\left(\frac{Q_0}{Q'}\right)^4}}
\right]\right)\!,\phantom{(99)} \\
\hspace*{-1em}
\label{Gqmod}
\Gamma_q(Q,Q';Q_0) &=& \frac{C_F}{2\pi} \frac{\alps(Q'^2)}{Q'^2} \left( \ln
\left(\frac{Q^2}{Q'^2}\right)-
\left[\frac{3}{2}
-3{\textstyle{\left(\frac{Q_0}{Q'}\right)}}
+\frac32{\textstyle{\left(\frac{Q_0}{Q'}\right)^2}}
\right]\right),
\end{eqnarray}
and all other results, (\ref{e9},\ref{e10},\ref{e2}--\ref{e8}),
remain unchanged.

We now have $n$-subjet rates that go smoothly to zero as $Y\to1$ for all
$n>1$ and smoothly to one for $n=1$.  However, the physical threshold
for $n$ massless partons to all be resolved from each other does not lie
at $Y=1$ but rather at $Y\equiv Y_n<1$.  We have not found a general
form for $Y_n$, but we explicitly find, for $n \le 4$, $Y_n=1/n^2$, and
$Y_5=4/(5+2\surd5)/5^2\approx0.422/5^2$.
It is worth noting that even exact numerical
solution of the original equation (\ref{master}) would not get the threshold
position right in general.  Fortuitously it does in our case for $n=2$,
but it does not, for example, in $e^+e^-$ annihilation, or in either
case for larger~$n$.

One, completely arbitrary, way to ensure that the $n$-subjet rate goes
smoothly to zero as $Y\to Y_n$ is to note that with logarithmic
accuracy, $Y$ can be replaced by any arbitrary rescaling
$Y(1+\cO(Y))$\cite{cat}.  For example, if we replace $Y$ by
$Y(1+\frac{1-Y_n}{Y_n^2}Y)$ then the $n$-subjet rates will go smoothly
to zero at $Y=Y_n$.  In particular, we replace
\begin{eqnarray}
  P_1(Y) &\longrightarrow& P_1\left(Y(1+12Y)\right), \\
  P_n(Y) &\longrightarrow& P_n\left(Y(1+n^2(n^2-1)Y)\right),
  \qquad 2 \le n \le 4.
\end{eqnarray}

\begin{figure}[ht]
\begin{minipage}[t]{0.475\textwidth}
\centerline{\resizebox{8cm}{!}{\rotatebox{90}{\includegraphics{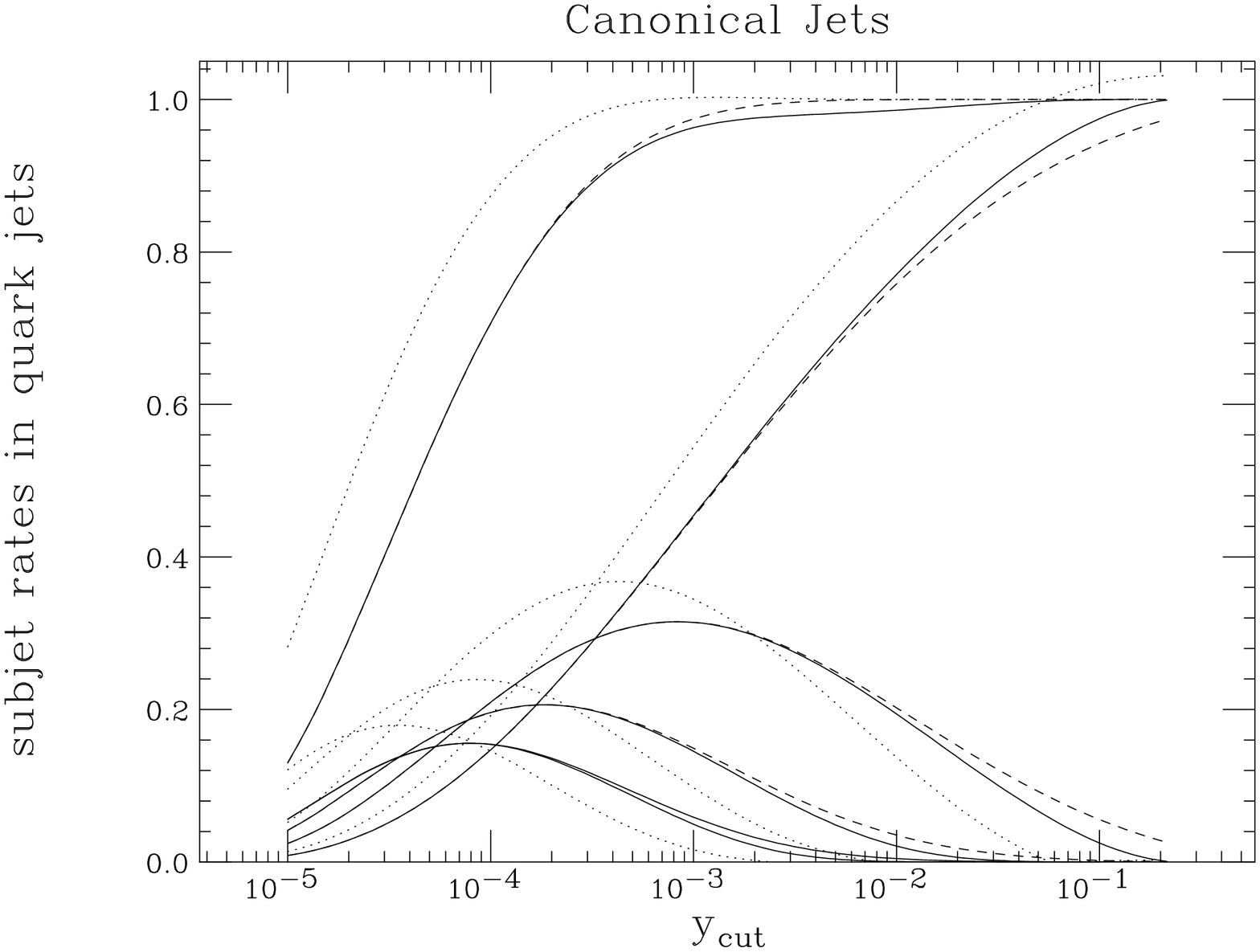}}}}
\caption{Quark subjet rates in the NLLA (dotted), with the inclusion of
$Q_0$-suppressed terms (dashed) and with shifted thresholds (solid). Also
shown is the sum of all subjet rates ($n \le 4$). Final state radiation only.}
\label{threshq}
\end{minipage}\hspace*{\fill}
\begin{minipage}[t]{0.475\textwidth}
\centerline{\resizebox{8cm}{!}{\rotatebox{90}{\includegraphics{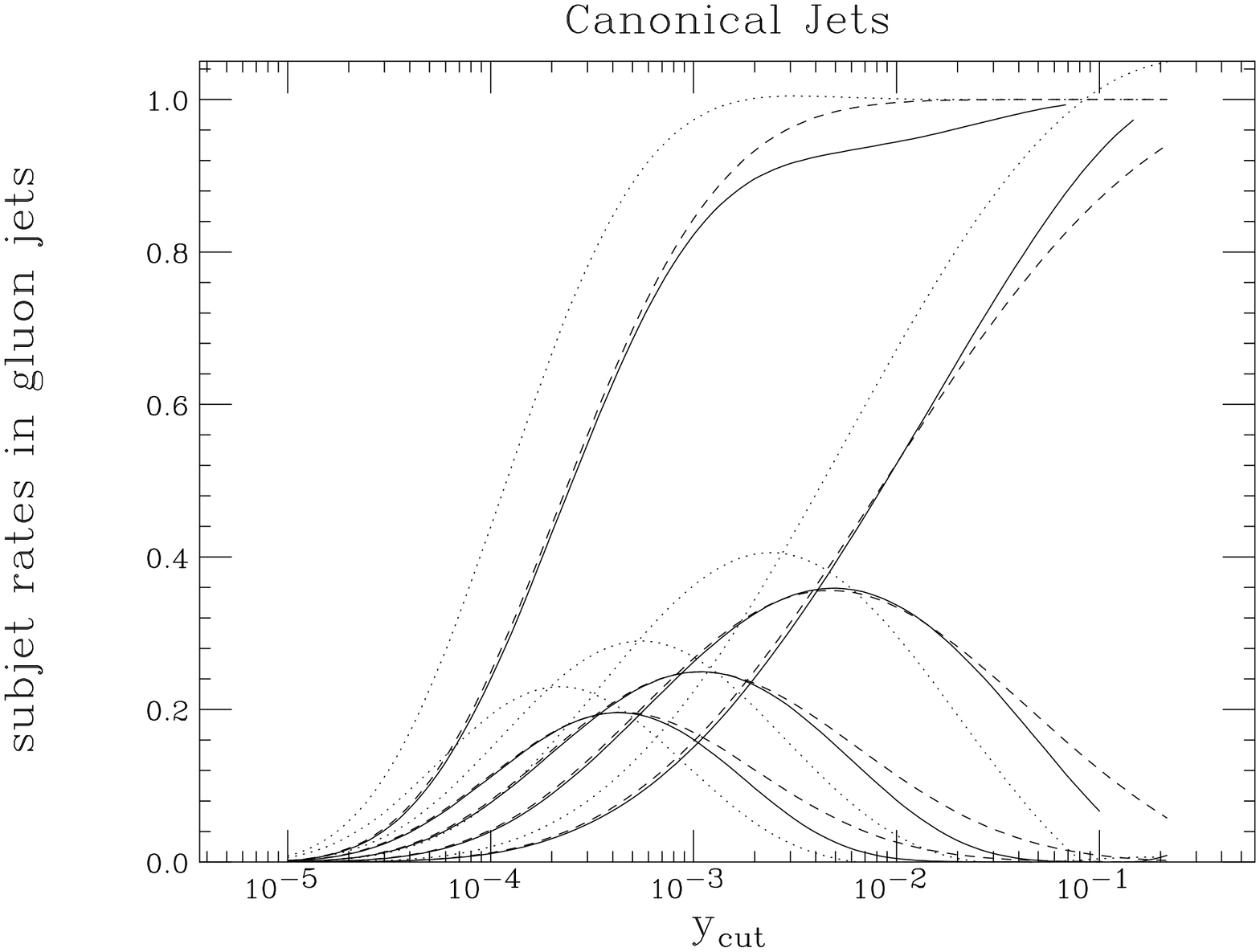}}}}
\caption{Gluon subjet rates in the NLLA (dotted), with the inclusion of
$Q_0$-suppressed terms (dashed) and with shifted thresholds (solid). Also
shown is the sum of all subjet rates ($n \le 4$). Final state radiation only.}
\label{threshg}
\end{minipage}
\end{figure}
We show results in Figures~\ref{threshq} and~\ref{threshg}.
They demonstrate that the threshold modifications make a considerable
difference to the $n$-subjet rates over quite a wide range in~$\ycut$.

Throughout this paper our canonical jets are defined to be produced in
$p\bar{p}$ collisions with $p_t = 60$~GeV, $\eta=0$, $R=1$ and $\surd s
= 1800$~GeV, and we use the CTEQ4M parton distribution functions~\cite{CTEQ}
as implemented in PDFLIB (\verb=NSET= = 34)~\cite{pdflib}.  In fact in
the final-state approximation we are discussing here, the properties of
quark and gluon jets depend only on their $p_t$ and the other parameters
are relevant only for determining the quark-to-gluon mix.

We will see later that, after matching with the leading order in
$\alps$ results, the modifications make less difference, although they are
certainly still not negligible. The suppression of the curves corresponding
to the sum of the first four subjet rates at
large $\ycut$ after including the threshold rescaling is indicative of the
potentially large corrections in the threshold region. We shall later see
that this suppression disappears after matching with the leading order in
$\alps$ results.

\subsection{Initial-state logarithms}
\label{inilogs}
An additional complication occurs in hadron collisions, because at
next-to-leading logarithmic accuracy gluons emitted by the incoming
partons can contribute to the number of subjets in a jet.  Since such
emission must be non-collinear one logarithm is lost, so to our
required accuracy it is sufficient to consider one initial-state gluon
emission, followed by the leading-logarithmic evolution of that gluon,
in which all subsequent emission is both soft and collinear to it.

We assume that the leading-order prediction for soft initial-state gluon
emission into a jet of flavour $a$ is known and given by
\begin{equation}
  \label{Aadef}
  dP_a = A_a\frac{d\kt^2}{\kt^2}\alps.
\end{equation}
The constant $A_a$ depends on the jet kinematics, the hadron collision
energy and type, the parton distribution functions, etc.  We return to
how it is calculated in the next section.  The jet's generating
function, (\ref{admixture}), then becomes
\begin{equation}
\GF(u,Q) = \sum_{a} \left[ F_a \G_a(u,Q) + \int_{Q_0^2}^{Q^2}
\frac{d\kt^2}{\kt^2} \alps(\kt^2) A_a \ \G_a(u,Q)(\G_g(u,\kt) -  1) \right].
\label{gen1}
\end{equation}
The first of the additional terms (the $\G_g(u,\kt)$ term) accounts for the
evolution of the soft initial-state gluon from the real emission.  The
second $(-1)$ term accounts for virtual corrections to the $2 \to 2$
matrix elements and is needed in order to conserve probability.

Since the non-collinear emission loses us one logarithm, we can work in
the LLA for the generating functions which appear in the second and
third terms on the right-hand-side of (\ref{gen1}).  The evolution
equation for the generating function of a parton can then simply be
obtained by truncating (\ref{master}) at LLA:
\begin{equation}
\frac{d \G_a(u,Q)}{d \ln Y} = - C_a \int_{Q_0^2}^{Q^2} \frac{d\kt^2}{\kt^2}
\frac{\alps(\kt^2)}{2\pi} \G_a(u,Q) ( \G_g(u,\kt) - 1).
\end{equation}
Hence we can rewrite (\ref{gen1}):
\begin{equation}
\GF(u,Q) = \sum_a \left\{ F_a \G_a(u,Q) -
\frac{2\pi A_a}{C_a} \frac{d \G_a(u,Q)}{d \ln Y} \right\}. \label{easy}
\end{equation}
In writing (\ref{easy}) we have avoided the need to perform the integral
over the transverse momentum of the soft gluon.  We just need to compute
$A_q$ and $A_g$.  Since doing this involves using the exact three-parton
matrix-element calculation, it gives us a very convenient way of
matching with the exact $\cO(\alps)$ result.

\subsection{Matching with exact \boldmath$\cO(\alps)$ result}
\label{fixed}
We begin by writing down the exact matrix-element result, before going
on to show how this can be used to calculate $A_a$ and hence match with
the resummed result.  The $\cO(\alps)$ result, which comes from the
ratio of the $\cO(\alps^3)$ three-parton and $\cO(\alps^2)$ two-parton
cross sections, is only non-zero for $P_1$ and $P_2$.  Furthermore, we
have the relation $P_1+P_2=1+\cO(\alps^2)$, which means that to this
order we only need to calculate $P_2$.  We define the $\cO(\alps)$ term
in the expansion of $P_2$ to be $R_2$ and obtain
\begin{equation}
R_2(\ycut,R;p_t,\eta,\phi) =
\frac{
\frac{(2 \pi)^4}{(2(2 \pi)^3)^3} p_t^3 \int_{-\etam}^{\etam} d \eta' \int_0^{2
  \pi} d \varphi \int_{\ycut}^{R^2/4} dy \int_{\sqrt{y}/R}^{1/2} dz
\frac{1-z}{2 z} \frac{1}{s \hat{s}} f_1(x_1) f_2(x_2) |\cM_3|^2 }{
\frac{(2 \pi)^4}{(2(2 \pi)^3)^2} p_t \int_{-\etam}^{\etam} d \eta'
\frac{1}{s \hat{s}} f_1(x_1) f_2(x_2) |\cM_2|^2 } \label{R2}
\end{equation}
in the $p_t$-scheme~\cite{MHS}.  The rescaled transverse momentum, $y$,
energy fraction, $z$ and azimuth around the jet axis, $\varphi$, are the
most convenient variables to describe the two-parton jet (along with
$p_t$, $\eta$ and $\phi$) and are integrated over the region where the
two partons are resolved as distinct subjets~\cite{MHS}.  It is
straightforward to perform these integrals numerically and hence compute
$R_2$.  We can also easily classify the jets into `quark',
i.e.~quark+gluon or antiquark+gluon, `gluon', i.e.~gluon+gluon or
quark+antiquark of the same flavour, or `other' contributions.  In all
plots we show the sum of all three contributions, together with the
separate quark and gluon contributions.

To this order, the generating function is therefore given by
\begin{equation}
\GF(u,Q) = u(1-R_2) + u^2 R_2 + \cO(\alps^2). \label{QED}
\end{equation}

In order to extract the leading-order probability of emitting a soft
initial-state gluon into the jet, we take $R_2$ and subtract from it the
$\cO(\alps)$ expansion of the final-state contribution:
\begin{eqnarray}
  \label{Pdef}
  P(\ycut,R;p_t,\eta,\phi) &=& F_q(p_t,\eta,\phi) P_q(\ycut,R;p_t,\eta,\phi)
+ F_g(p_t,\eta,\phi) P_g(\ycut,R;p_t,\eta,\phi) \nonumber \\
& = & R_2(\ycut,R;p_t,\eta,\phi) - \Delta_2(\ycut/R^2;p_t,\eta,\phi),
\end{eqnarray}
where
\begin{equation}
\Delta_2(Y;p_t,\eta,\phi) = F_q(p_t,\eta,\phi)\Delta_{2q}(Y)
                          + F_g(p_t,\eta,\phi)\Delta_{2g}(Y),
\end{equation}
\begin{eqnarray}
\label{D2qdef}
\Delta_{2q}(Y) &=& \int_{Q_0^2}^{Q^2} dQ'^2 \, \Gamma_q(Q,Q') +
\cO(\alps^2)
\\&=& \frac{\alps}{4 \pi} (C_F \ln^2 Y + 3 C_F \ln Y), \\
\label{D2gdef}
\Delta_{2g}(Y) &=& \int_{Q_0^2}^{Q^2} dQ'^2 \, (\Gamma_g(Q,Q')+\Gamma_f(Q'))
+ \cO(\alps^2)
\\&=& \frac{\alps}{4 \pi} (C_A \ln^2 Y + \beta_0 \ln Y).
\end{eqnarray}
The only remaining logarithmically-enhanced terms must come from
initial-state radiation.

According to the definition of $A_a$, (\ref{Aadef}), the
logarithmically-enhanced part of $P$ is also equal to
\begin{equation}
  \label{P}
  P_a  = \alps A_a \ln(1/Y) + \cO(\alps),
\end{equation}
where the neglected terms are pure $\cO(\alps)$ with no logarithmic
enhancement.  We can therefore equate (\ref{Pdef}) and (\ref{P}) to
logarithmic accuracy, to give
\begin{equation}
  A_a= \frac{(R_2 - \Delta_2)_a}{\alps(Q^2)\ln Y}.
\end{equation}
Inserting this into (\ref{easy}), we arrive at our central result
for the generating function of jets in hadron collisions at scale
$Q=p_t$,
\begin{equation}
\GF(u,Q) = \sum_a \left[ F_a \G_a(u,Q)
+ \frac{2 \pi (R_2 - \Delta_2)_a}{C_a \alps(Q^2)\ln Y}\;
\frac{d \G_a(u,Q)}{d \ln Y} \right] -u R_2^{\mathrm{others}} +
u^2 R_2^{\mathrm{others}}.
\label{central}
\end{equation}
We recall that $\G_a(u,Q)$ are computed in the NLLA as given in
(\ref{e9},\ref{e10},\ref{e2}--\ref{e8}) and that $d\G_a(u,Q)/d\ln Y$
are computed in the LLA, by using the same formul\ae, but keeping only
the logarithmically-enhanced terms in $\Gamma_a$,
(\ref{e12},\ref{e13}).

Note that by computing $R_2$ using the full $2 \to 3$ matrix elements,
we are going beyond the NLLA\@.  This is because we are guaranteed to
compute the lowest order in $\alps$ contribution exactly, i.e.~we have
succeeded in matching the NLL approach with the fixed order calculation.
We have therefore attained the accuracy aimed for in (\ref{accuracy}).
The terms involving $R_2^{\mathrm{others}}$ in (\ref{central}) are $\cO(\alps)$
terms that lie outside of the NLLA but are needed to formally
ensure the matching. They are responsible for describing jets that are
neither quark jets nor gluon jets and, as mentioned earlier, they are
negligibly small.

We finally note that when using the modified results of
Section~\ref{thresh}, the modifications affect not only the resummed parts
of (\ref{central}), but also the matching term $\Delta_2$.  Since
this subtracts off the double-counting between the fixed-order term
$R_2$ and the resummed results, and $R_2$ already contains the correct
threshold behaviour, $\Delta_2$ must be calculated in the same way as
the resummed results.  Specifically, the inclusion of the
$Q_0$-suppressed terms in the evolution affects $\G_a$ and $\Delta_2$
through equations (\ref{D2qdef},\ref{D2gdef}) and (\ref{Gfmod}--\ref{Gqmod})
and the rescaling of $Y$ affects all results except $R_2$.

\begin{figure}[ht]
\begin{minipage}[t]{0.475\textwidth}
\centerline{\resizebox{8cm}{!}{\rotatebox{90}{\includegraphics{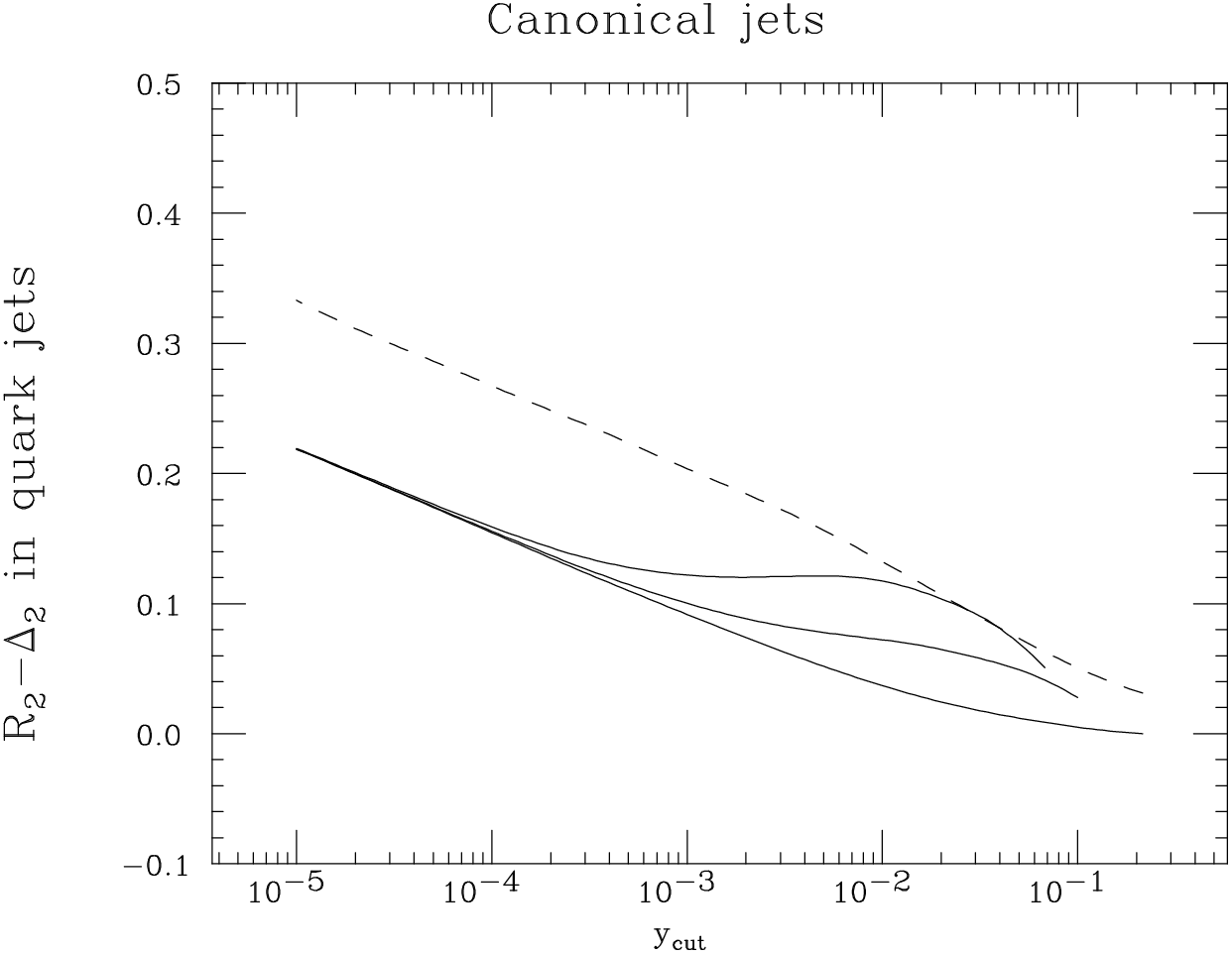}}}}
\caption{Comparison of $(R_2-\Delta_2)_q$ computed in the NLLA (dashed) with
that computed after including threshold modifications (solid).}
\label{r2dq}
\end{minipage}\hspace*{\fill}
\begin{minipage}[t]{0.475\textwidth}
\centerline{\resizebox{8cm}{!}{\rotatebox{90}{\includegraphics{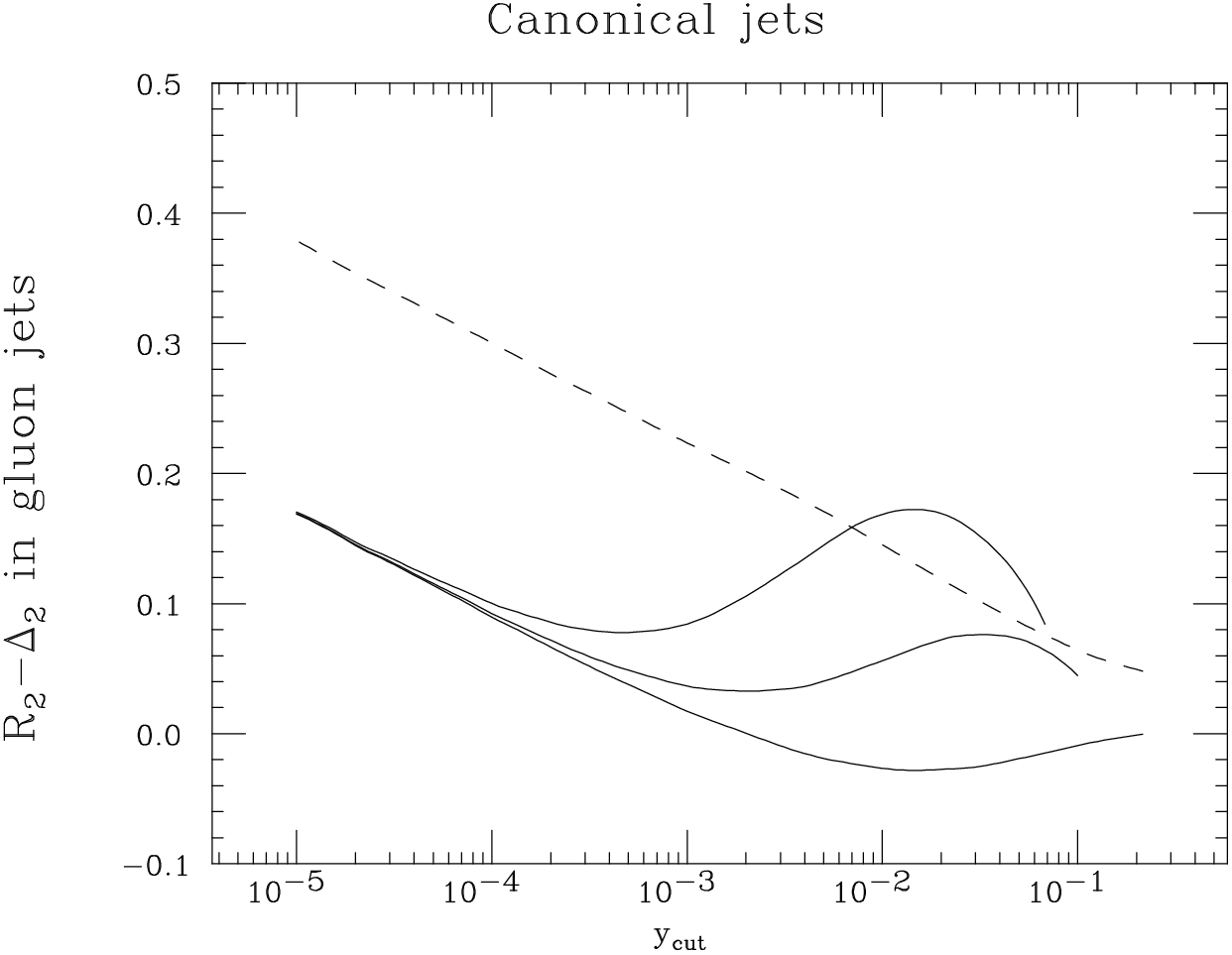}}}}
\caption{Comparison of $(R_2-\Delta_2)_g$ computed in the NLLA (dashed) with
that computed after including threshold modifications (solid).}
\label{r2dg}
\end{minipage}
\end{figure}
In Figures~\ref{r2dq} and~\ref{r2dg} we show $R_2 - \Delta_2$ for quark
and gluon jets, without and with the threshold modifications.  Since the
latter are different for the different subjet rates, there are three
solid curves corresponding to the one/two subjet rates (lower curve),
three subjet rate (middle curve) and four subjet rate (upper curve).  It
can be clearly seen that for small $Y$, $R_2 - \Delta_2$ consists of a NLL
piece ($\sim \alps \ln Y$) and a fixed order piece ($\sim \alps$), i.e.
\begin{equation}
R_2 - \Delta_2 \sim \alps (~\ln Y + \mathrm{constant}~).
\label{little}
\end{equation}
Physically the logarithmically-rising term corresponds to initial-state
radiation, while the constant term cannot uniquely be ascribed to the
initial or final state.  Note that the coefficient of the logarithm is
unchanged by the threshold matching, while the constant term, which is
small in the NLLA, becomes significant throughout the $\ycut$ region we
consider.  We stress that this non-logarithmic part is included to
ensure proper matching with the lowest order matrix element.  We have
chosen to multiply it by the LLA parton multiplication factor,
$d\G_a(u,Q)/d\ln Y$.  In so doing we have introduced a tower of
sub-leading terms that lie beyond our level of approximation.

Looking at the structure of (\ref{central}), it is tempting to think of
the first term as arising solely from final-state radiation and the
second solely from initial-state radiation.  Indeed this is the case for
logarithmically-enhanced terms, but constant terms like the one in
(\ref{little}) are not uniquely part of either contribution.  The fact
that $R_2-\Delta_2$ is small at some $\ycut$ value should not therefore be
assumed to mean that initial-state effects are small there.

Note that the slopes of Figures~\ref{r2dq} and~\ref{r2dg}, i.e.~the
coefficients of the initial-state logarithms for quark and gluon jets,
are rather similar.  In fact in \cite{MHS}, it was mentioned that in the
planar approximation (i.e.~in the limit of large number of
colours,~$N_c$) they would be the same.  Their numerical difference
(about~20\%) is consistent with this statement.  Moreover it was found
that the slope was largely independent of their production mechanism,
\emph{provided\/} the rapidity of the other jet in the event was
integrated out.  If the recoiling jet's rapidity is constrained, the
slope of $R_2-\Delta_2$ becomes dependent on it.  We return to this point
later.

It is also worth pointing out that the initial-state component is
roughly proportional to~$R^2$, so reducing $R$ can make the relative
effect of initial-state radiation considerably smaller\cite{DZ}.

\section{Numerical results}

We recall that our canonical jets are defined to be produced in $p\bar{p}$
collisions with $p_t = 60$~GeV, $\eta=0$, $R=1$ and $\surd s = 1800$~GeV,
and we use the CTEQ4M parton distribution functions~\cite{CTEQ} as implemented
in PDFLIB (\verb=NSET= = 34)~\cite{pdflib}.

\begin{figure}[ht]
\begin{minipage}[t]{0.475\textwidth}
\centerline{\resizebox{8cm}{!}{\rotatebox{90}{\includegraphics{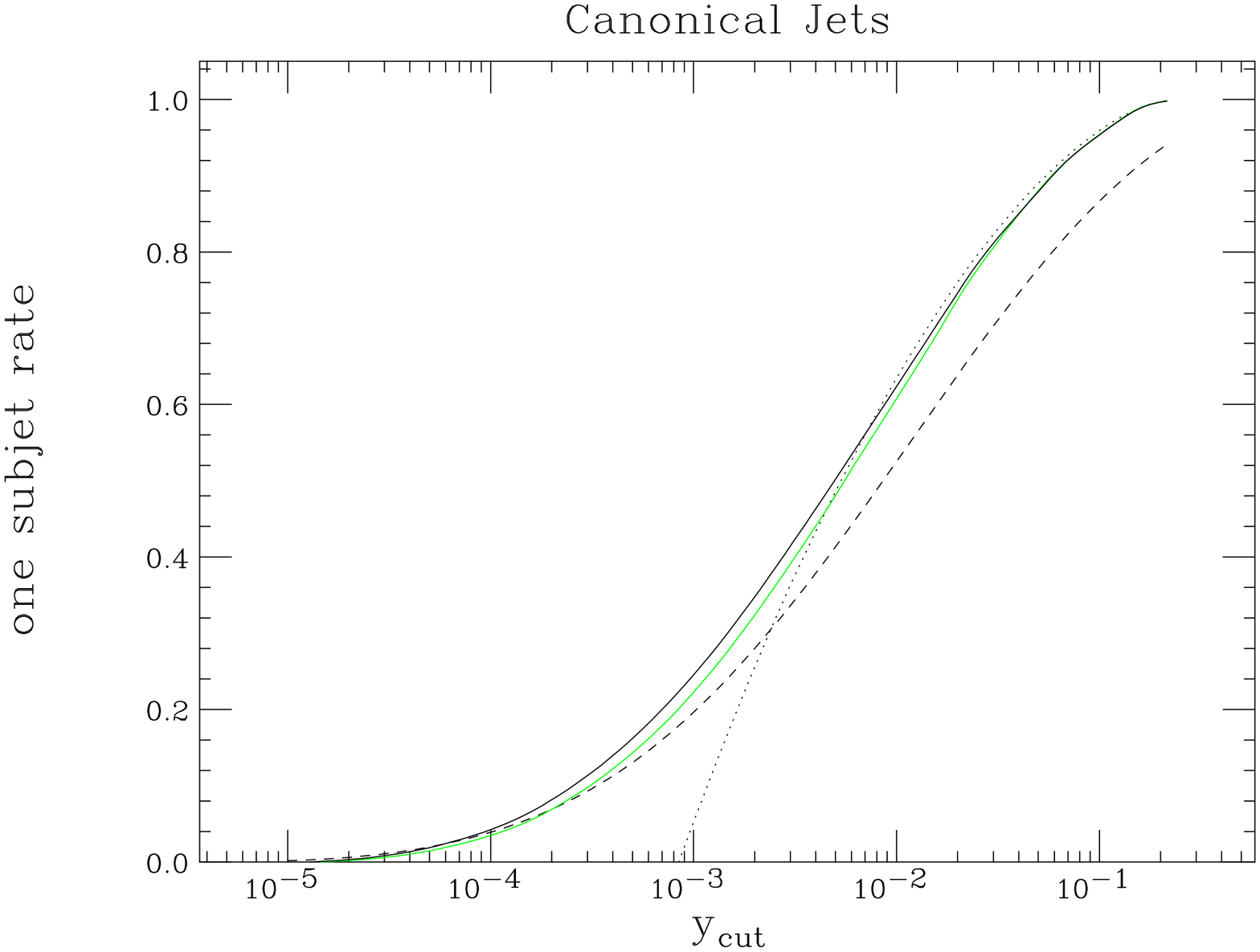}}}}
\caption{One subjet rate: comparison of leading order (dotted), LLA (dashed),
matched NLLA (solid) and matched NLLA with threshold modifications (green).}
\label{compare1}
\end{minipage}\hspace*{\fill}
\begin{minipage}[t]{0.475\textwidth}
\centerline{\resizebox{8cm}{!}{\rotatebox{90}{\includegraphics{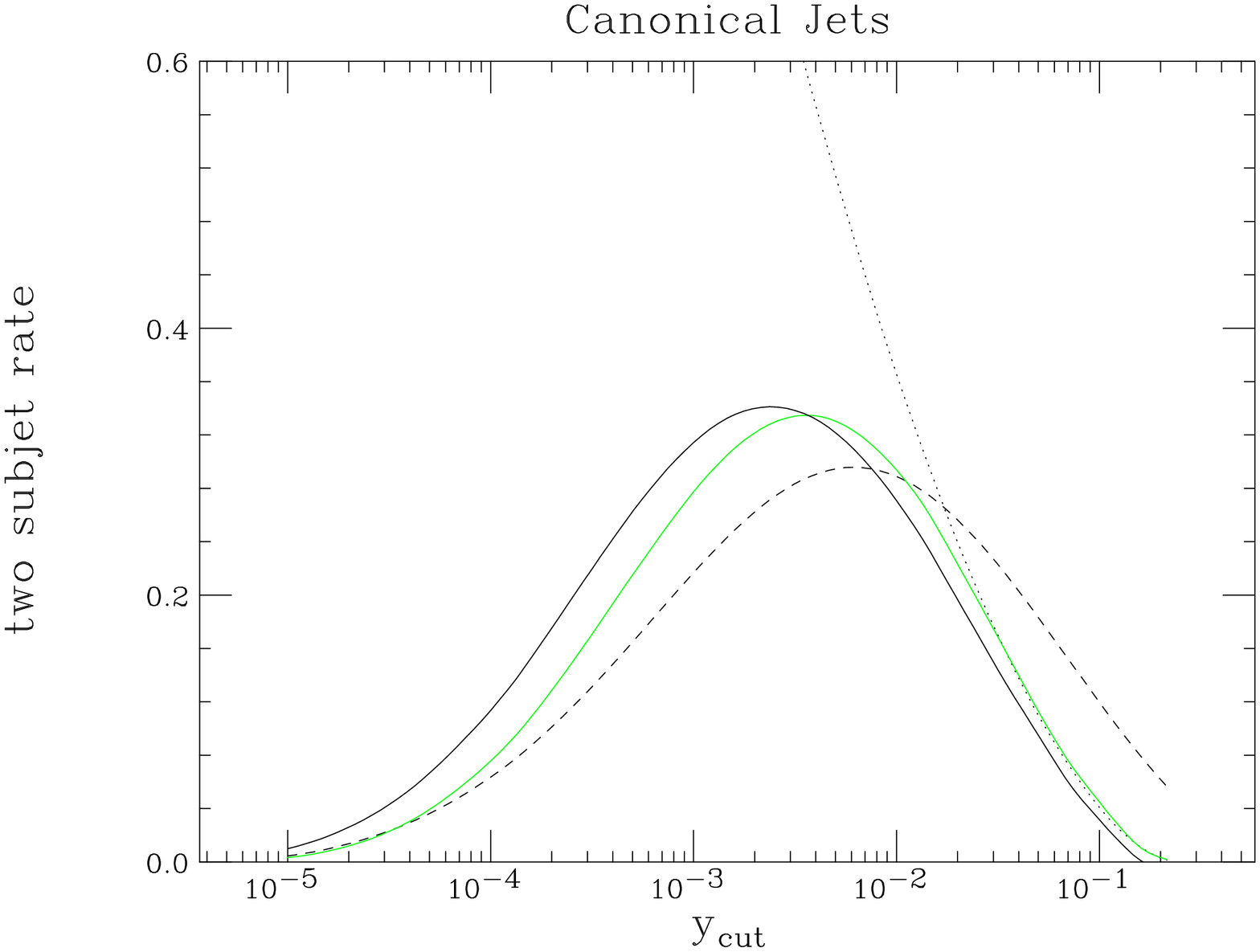}}}}
\caption{Two subjet rate: comparison of leading order (dotted), LLA (dashed),
matched NLLA (solid) and matched NLLA with threshold modifications (green).}
\label{compare2}
\end{minipage}
\end{figure}
\begin{figure}[ht]
\begin{minipage}[t]{0.475\textwidth}
\centerline{\resizebox{8cm}{!}{\rotatebox{90}{\includegraphics{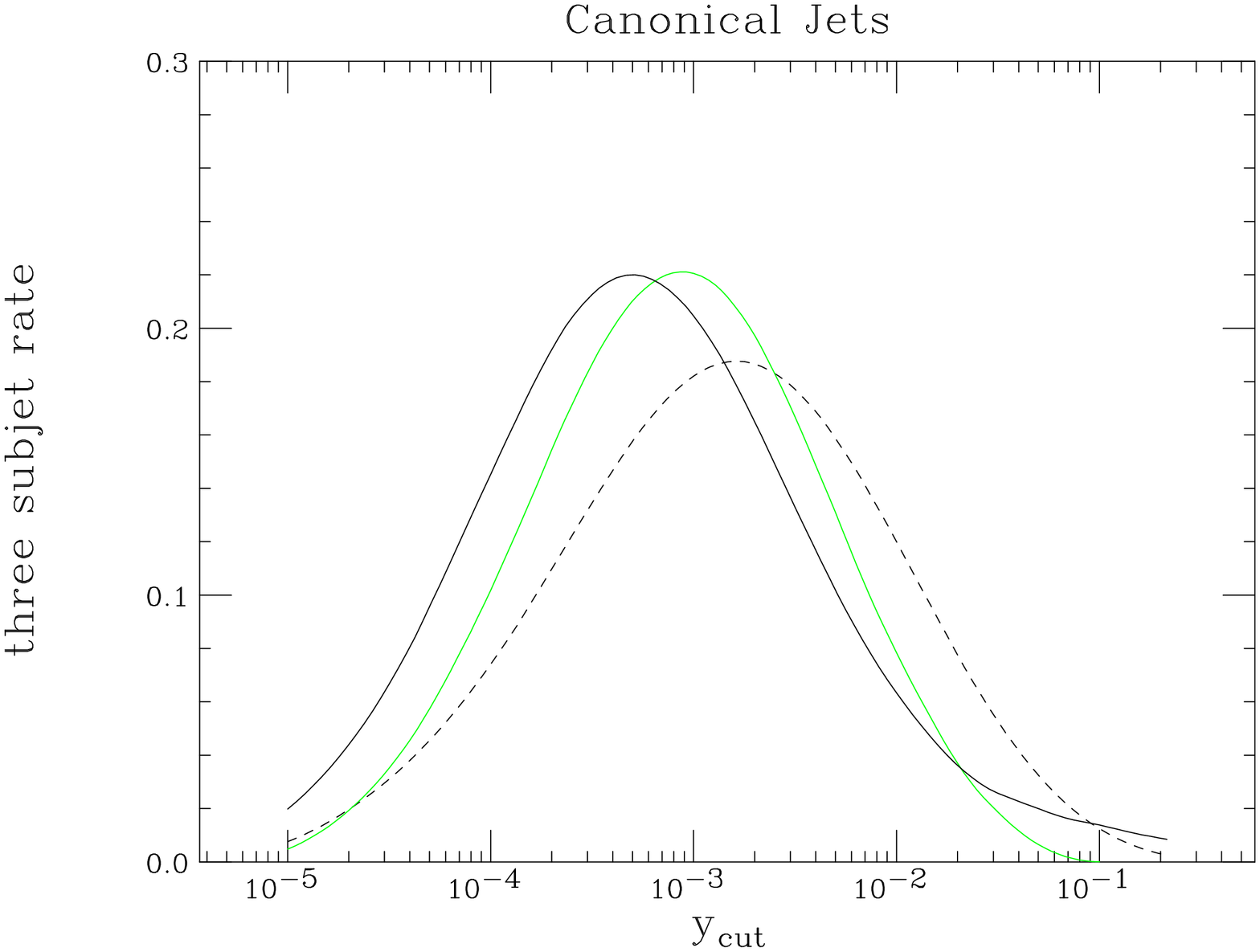}}}}
\caption{Three subjet rate: comparison of LLA (dashed),
matched NLLA (solid) and matched NLLA with threshold modifications (green).}
\label{compare3}
\end{minipage}\hspace*{\fill}
\begin{minipage}[t]{0.475\textwidth}
\centerline{\resizebox{8cm}{!}{\rotatebox{90}{\includegraphics{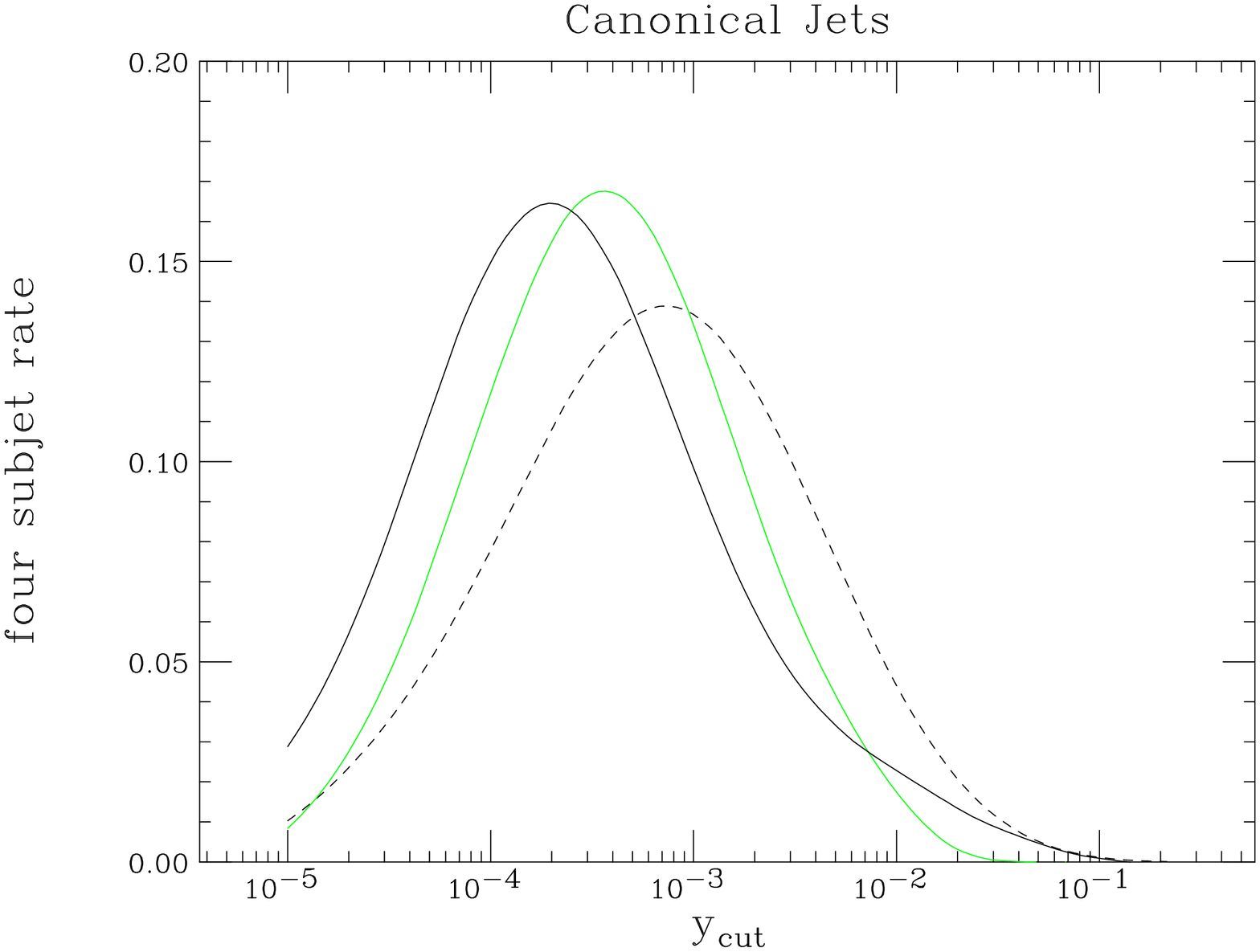}}}}
\caption{Four subjet rate: comparison of LLA (dashed),
matched NLLA (solid) and matched NLLA with threshold modifications (green).}
\label{compare4}
\end{minipage}
\end{figure}
Figures~\ref{compare1}--\ref{compare4} show the individual subjet rates in
canonical jets for up to and including four subjets.
Comparison is made between the leading order prediction, the prediction
to leading logarithmic accuracy (LLA) and the fully matched
next-to-leading logarithmic accuracy prediction (NLLA) with and without
the improved treatment of the threshold region discussed in
Sect.~\ref{thresh}.
Firstly, we see that the fixed order results have only a small
region of reliability, rapidly becoming unphysical for $\ycut<10^{-2}$.
The LLA result is physically behaved, but the higher subjet rates grow much
too quickly away from threshold.  The matched NLLA results on the other
hand should be reliable for all $\ycut$.  They approximate the fixed-order
results for $P_1$ and $P_2$ for large $\ycut$ and remain well-behaved
for small $\ycut$.  We see that the modified threshold treatment makes
very little difference for $P_1$, slightly more for $P_2$ and more still
for the higher subjet rates.  This is an indication of the relative
importance of neglected next-to-next-to-leading logarithmic effects and
hence of the accuracy of the whole calculation.  This difference would
be reduced significantly by matching to the relevant $n$-parton
tree-level matrix elements, probably to a similar level to $P_2$, but to
reduce the dependence further would require working to at least NNLLA\@.
For all subsequent figures we use the threshold-improved results.

\begin{figure}[ht]
\begin{minipage}[t]{0.475\textwidth}
\centerline{\resizebox{8cm}{!}{\rotatebox{90}{\includegraphics{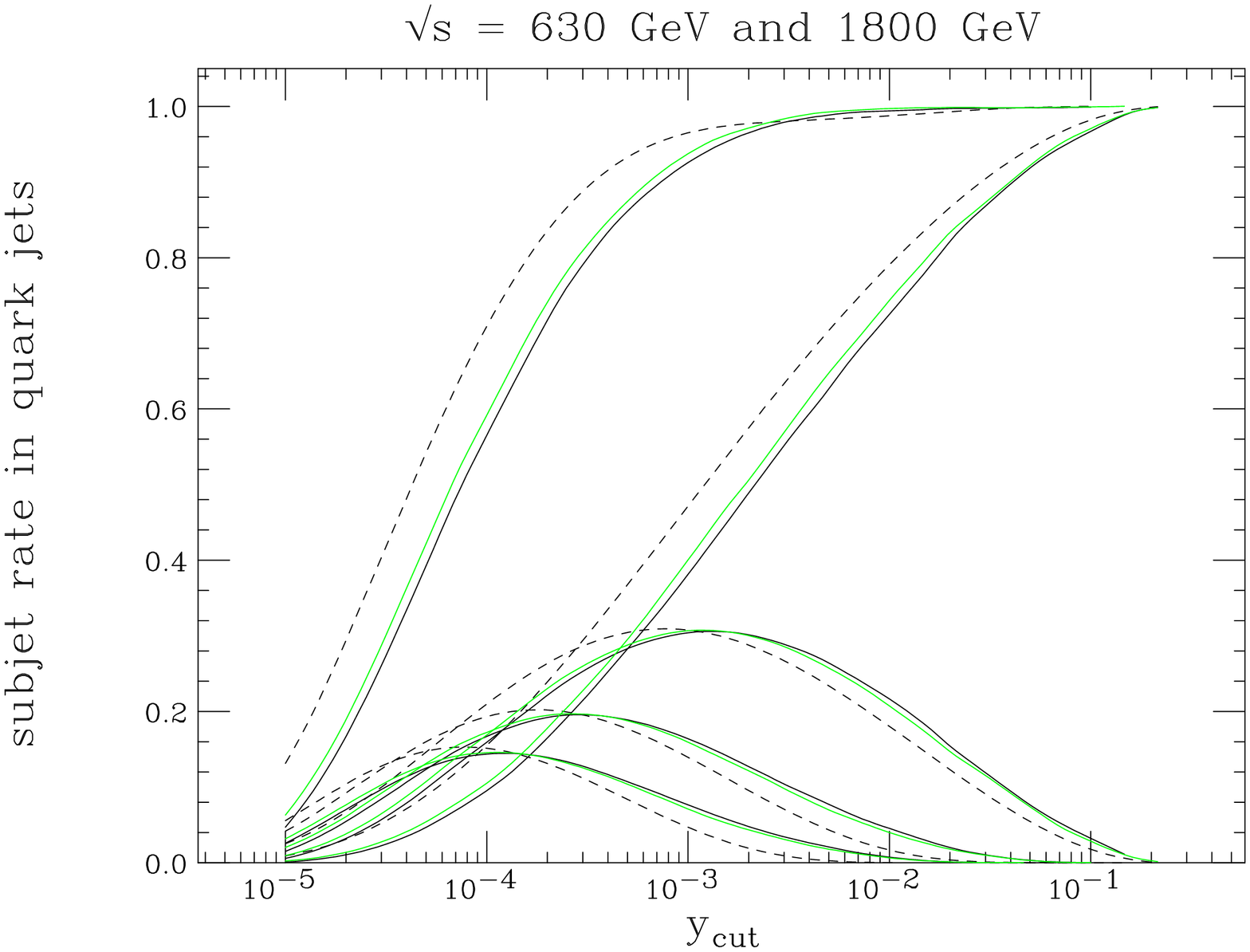}}}}
\caption{$\ycut$ dependence of subjet rates in quark jets at 630 GeV (green)
and 1800 GeV (black) and from final-state logarithms only (dashed).
Also shown is the sum of all the rates.}
\label{quark1}
\end{minipage}\hspace*{\fill}
\begin{minipage}[t]{0.475\textwidth}
\centerline{\resizebox{8cm}{!}{\rotatebox{90}{\includegraphics{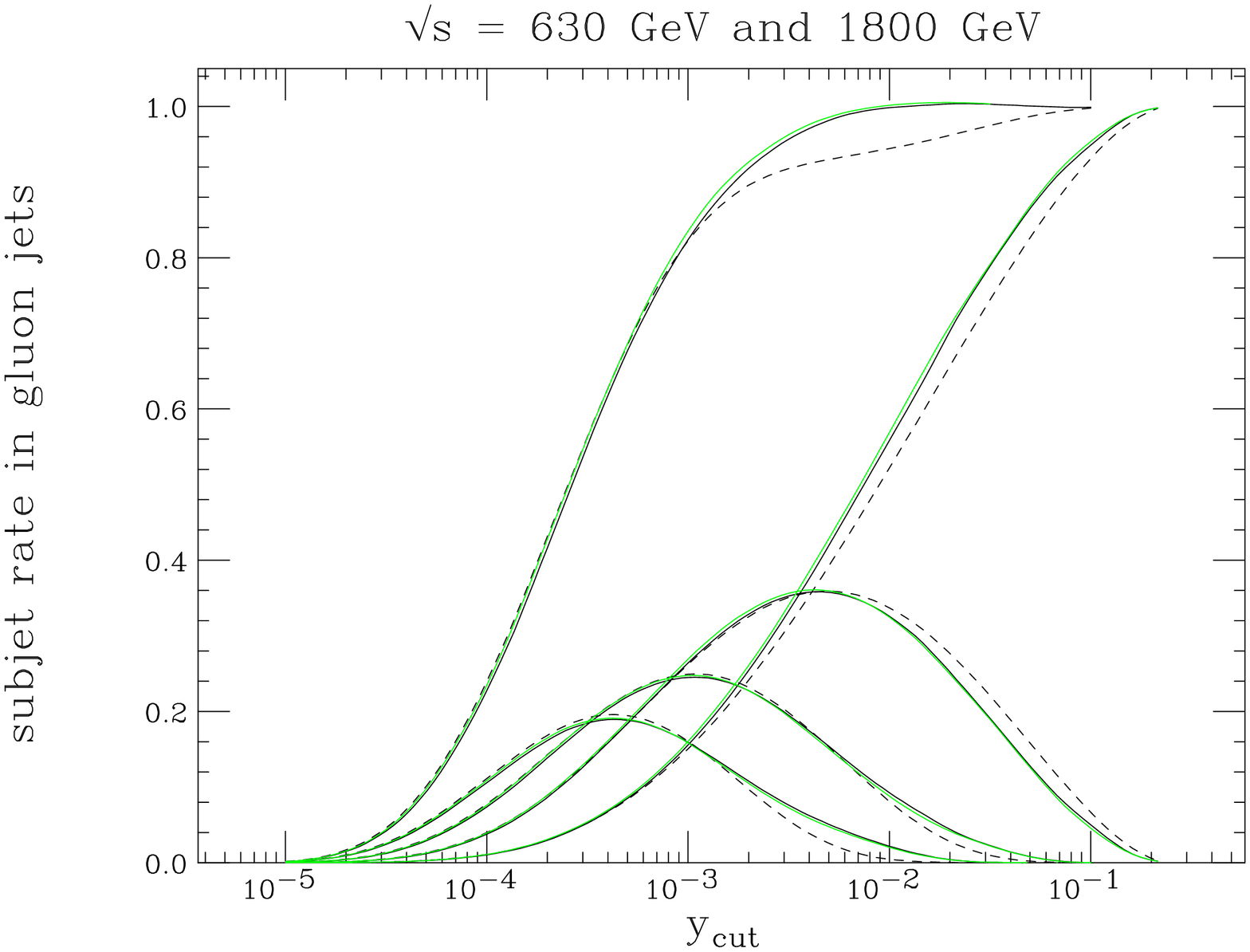}}}}
\caption{$\ycut$ dependence of subjet rates in gluon jets at 630 GeV (green)
and 1800 GeV (black) and from final-state logarithms only (dashed).
Also shown is the sum of all the rates.}
\label{gluon1}
\end{minipage}
\end{figure}
\begin{figure}[ht]
\begin{minipage}[t]{0.475\textwidth}
\centerline{\resizebox{8cm}{!}{\rotatebox{90}{\includegraphics{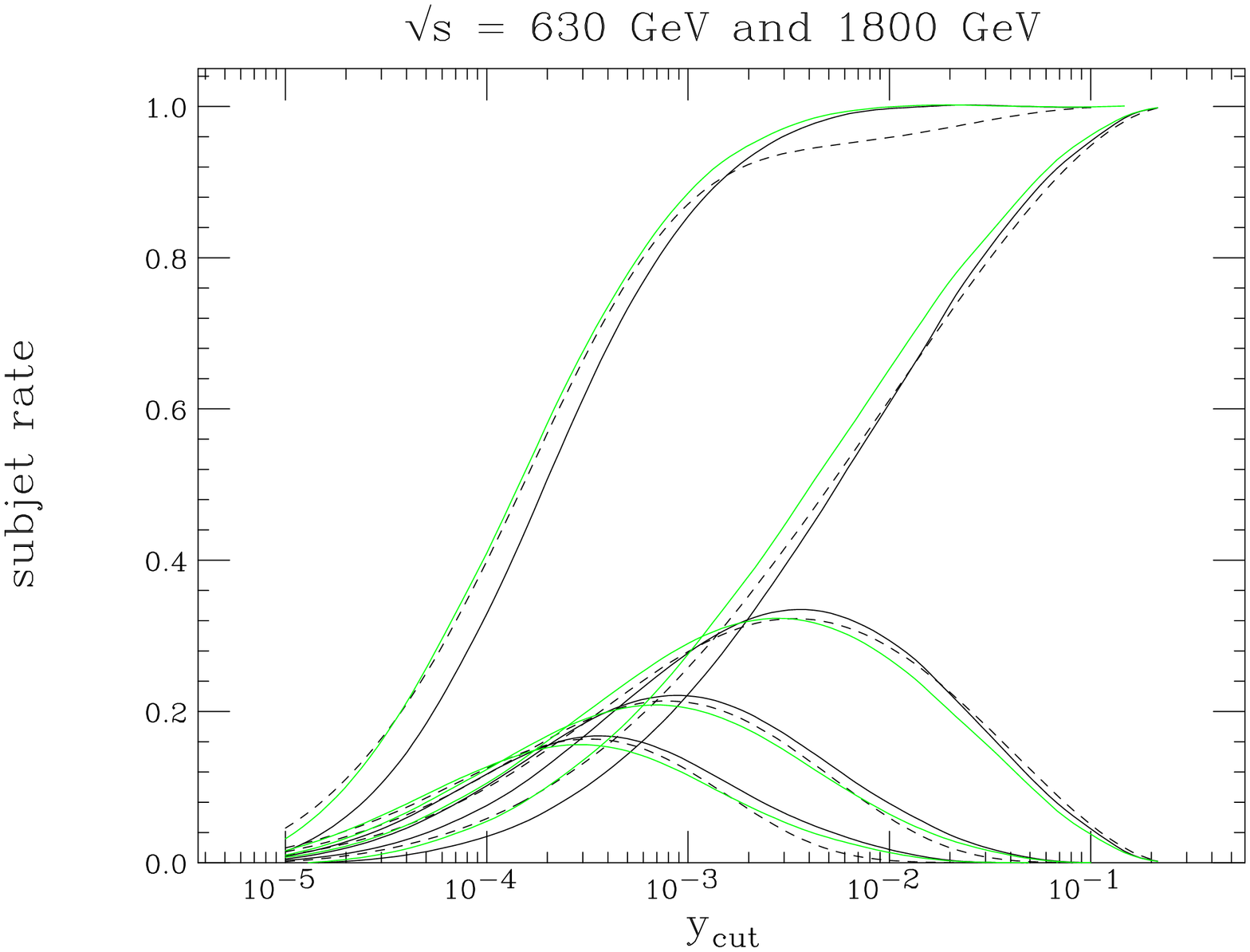}}}}
\caption{$\ycut$ dependence of subjet rates in all jets at 630 GeV (green)
and 1800 GeV (black) and from final-state logarithms only (dashed).
Also shown is the sum of all the rates.}
\label{all1}
\end{minipage}\hspace*{\fill}
\begin{minipage}[t]{0.475\textwidth}
\centerline{\resizebox{8cm}{!}{\rotatebox{90}{\includegraphics{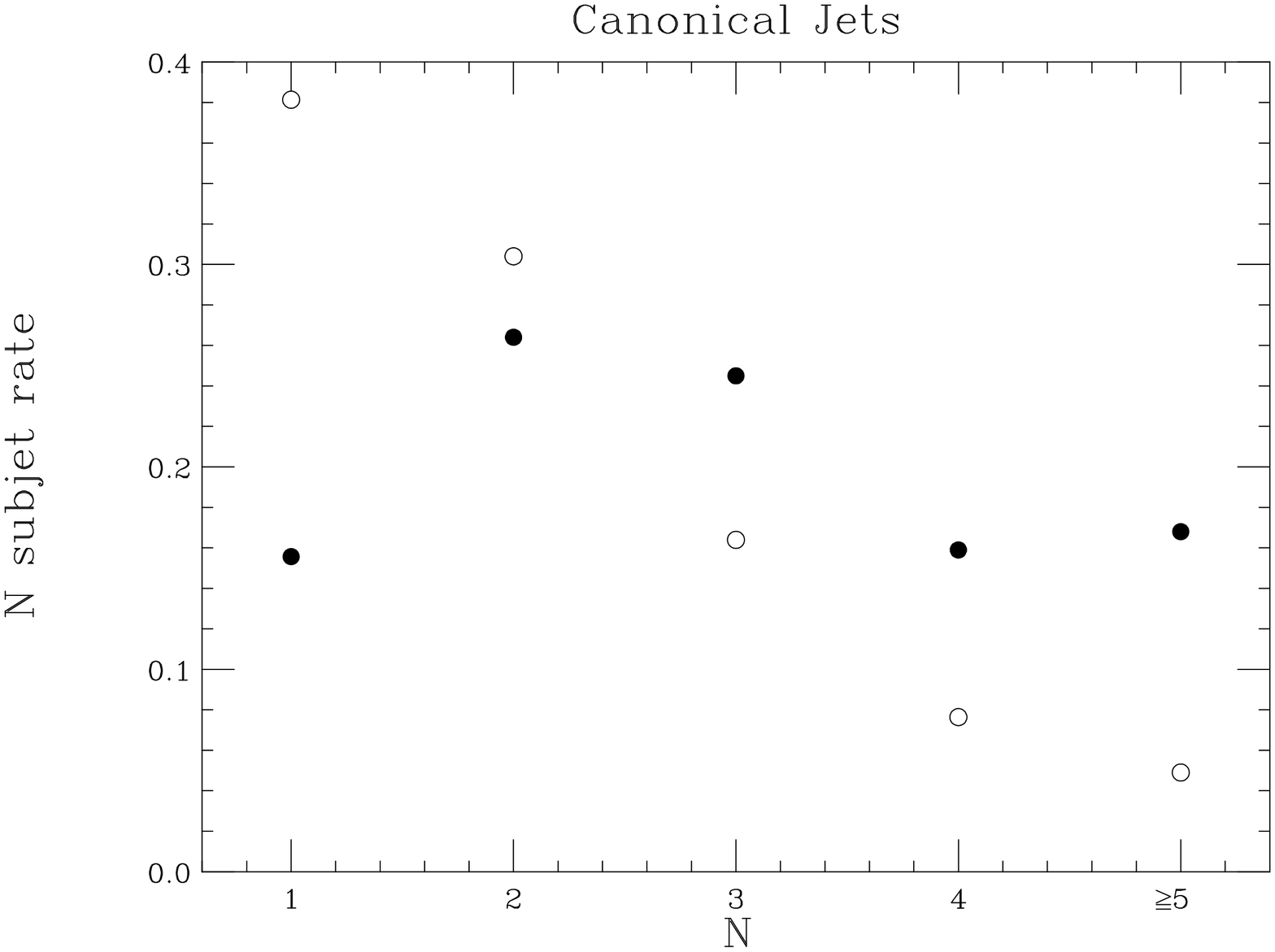}}}}
\caption{Rates for $N$ subjets at $\ycut~=~10^{-3}$ in quark (open circles)
and gluon (solid circles) jets. The points in the $N=5$ bin are for $N \ge 5$
subjets.}
\label{QGN}
\end{minipage}
\end{figure}
In Figure~\ref{quark1} we show the $\ycut$ dependence of the individual
subjet rates in quark jets at the two different collider energies of 630 GeV
and 1800 GeV.  Figure~\ref{gluon1} is a similar plot but for gluon jets and
Figure~\ref{all1} is for all jets.
Note the very weak dependence of the rates in quark and gluon jets on
the centre-of-mass energy.  This result supports the recent D\O\ analysis where
it is assumed that jet observables do not depend upon centre-of-mass
energy~\cite{DZ}.
One might be tempted to assume, seeing this result and the later ones,
that because the properties of quark and gluon jets depend so little on
how they were produced they are dominated by final state effects.
Comparison with the final-state-only curves in Figures~\ref{quark1}
and~\ref{gluon1} shows that this is not the case.  A significant
fraction of the resolved subjets come from non-final-state radiation but
are nevertheless still universal to a very good approximation.  The
all-jets results do vary with centre-of-mass energy, because the
quark-to-gluon mix is varying

As discussed at the end of Section~\ref{fixed}, the effect of initial
state radiation cannot be inferred simply from the difference between
the dashed and solid lines in Figures~\ref{quark1} and~\ref{gluon1}.
The difference represents the full effect of the $R_2-\Delta_2$ term in
(\ref{central}), after applying the threshold corrections.  This
contribution is made up of a logarithmic piece and an order $\alps$
piece (see~(\ref{little})).  The logarithmic piece can be termed initial
state radiation whilst the latter cannot.

Figure~\ref{QGN} shows the individual subjet rates at fixed $\ycut=10^{-3}$ for
canonical quark and gluon jets.  The point at $N=5$ is the inferred rate for
5 or more subjets. It has been computed using $P_{n \ge 5}^a = 1 -
\sum_{i=1}^4 P_i^a,$ with the first four subjet rates all shifted to
have thresholds at $Y_5$, so that $P_{n \ge 5}^a$ is sensibly-behaved
there.

\begin{figure}[ht]
\begin{minipage}[t]{0.475\textwidth}
\centerline{\resizebox{8cm}{!}{\rotatebox{90}{\includegraphics{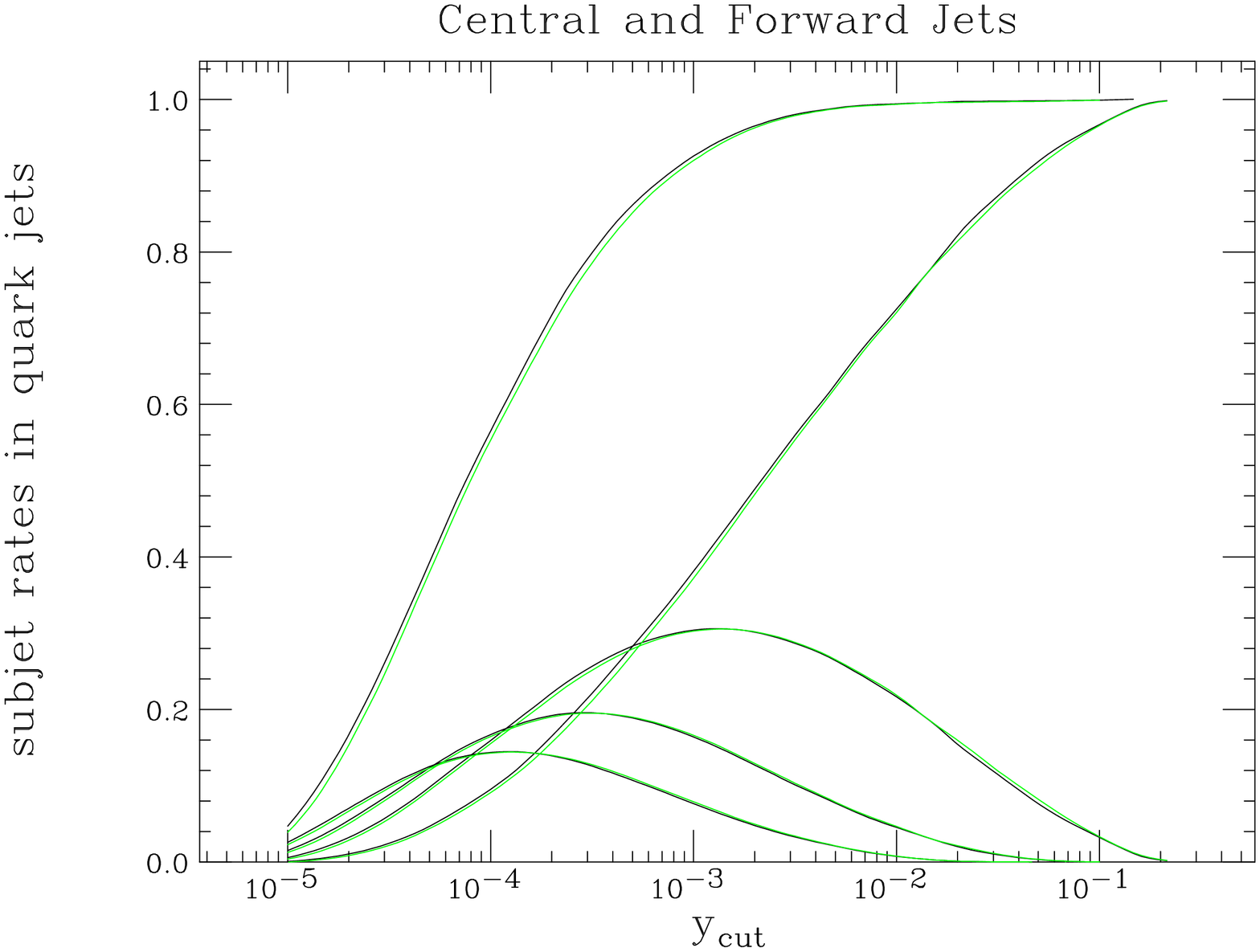}}}}
\caption{Comparison of subjet rates for $\eta=0$ (black) and
$\eta=2$ (green) quark jets.}
\label{etaq}
\end{minipage}\hspace*{\fill}
\begin{minipage}[t]{0.475\textwidth}
\centerline{\resizebox{8cm}{!}{\rotatebox{90}{\includegraphics{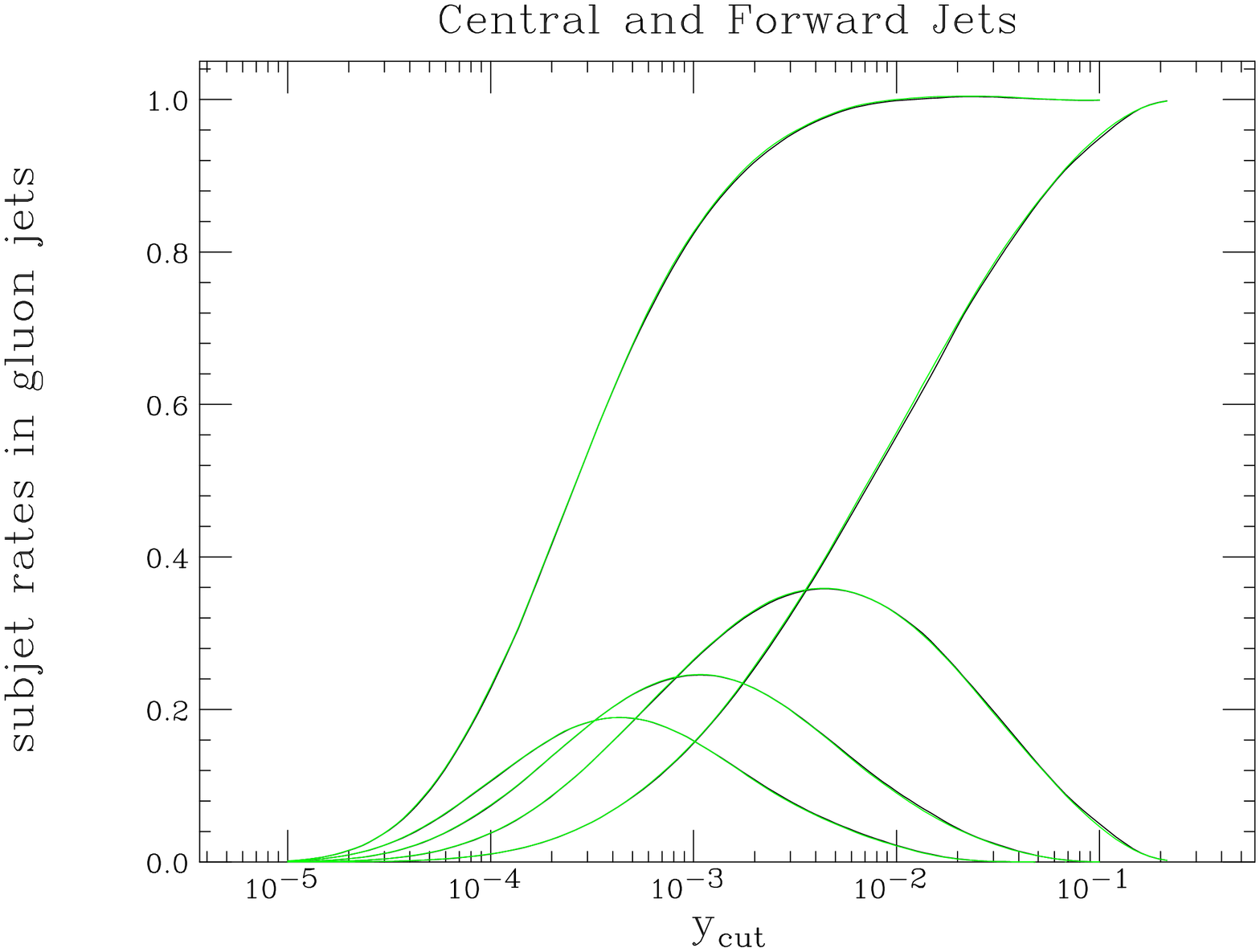}}}}
\caption{Comparison of subjet rates for $\eta=0$ (black) and
$\eta=2$ (green) gluon jets.}
\label{etag}
\end{minipage}
\end{figure}
\begin{figure}[ht]
\begin{minipage}[t]{0.475\textwidth}
\centerline{\resizebox{8cm}{!}{\rotatebox{90}{\includegraphics{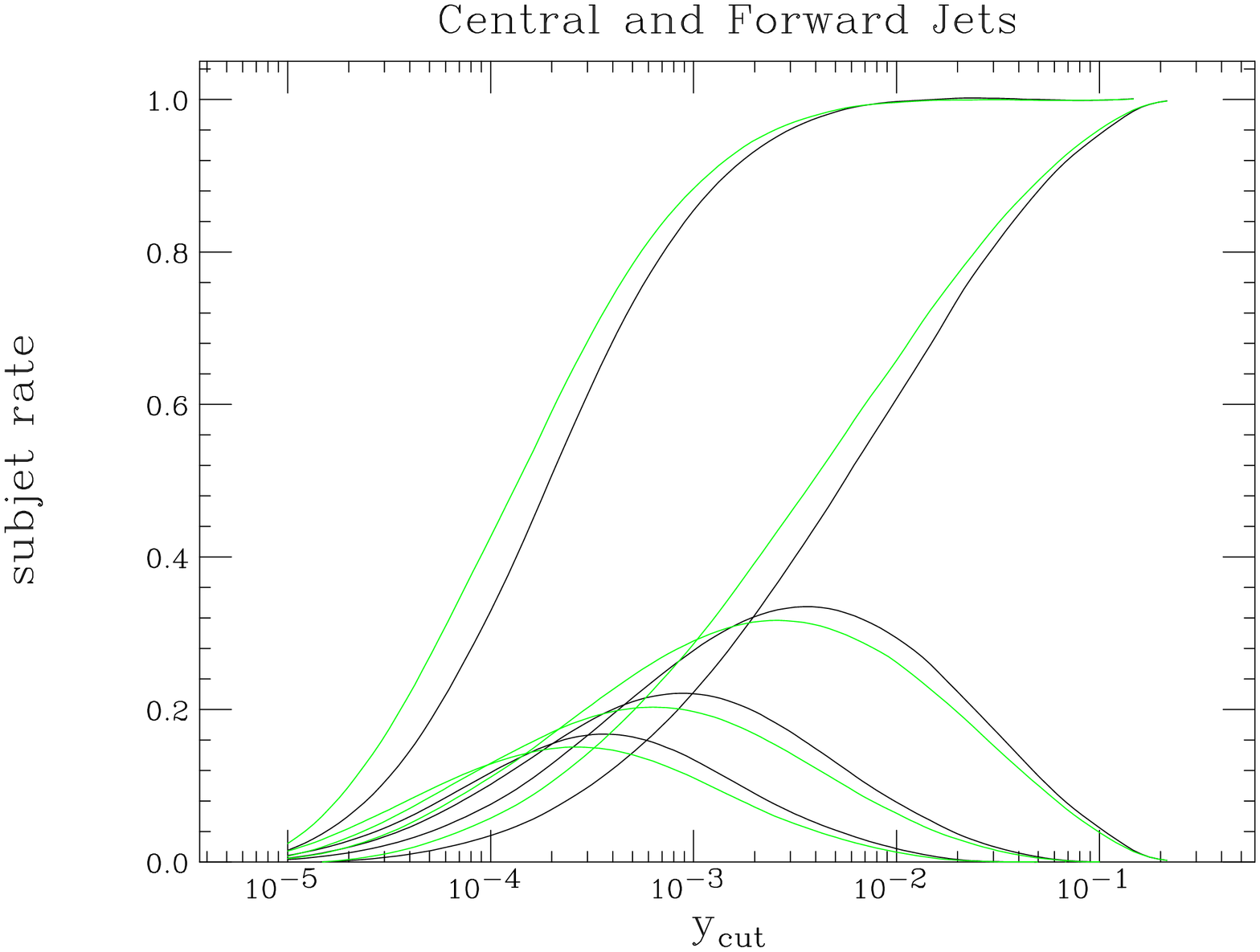}}}}
\caption{Comparison of subjet rates for $\eta=0$ (black) and
$\eta=2$ (green) jets.}
\label{eta}
\end{minipage}\hspace*{\fill}
\begin{minipage}[t]{0.475\textwidth}
\centerline{\resizebox{8cm}{!}{\rotatebox{90}{\includegraphics{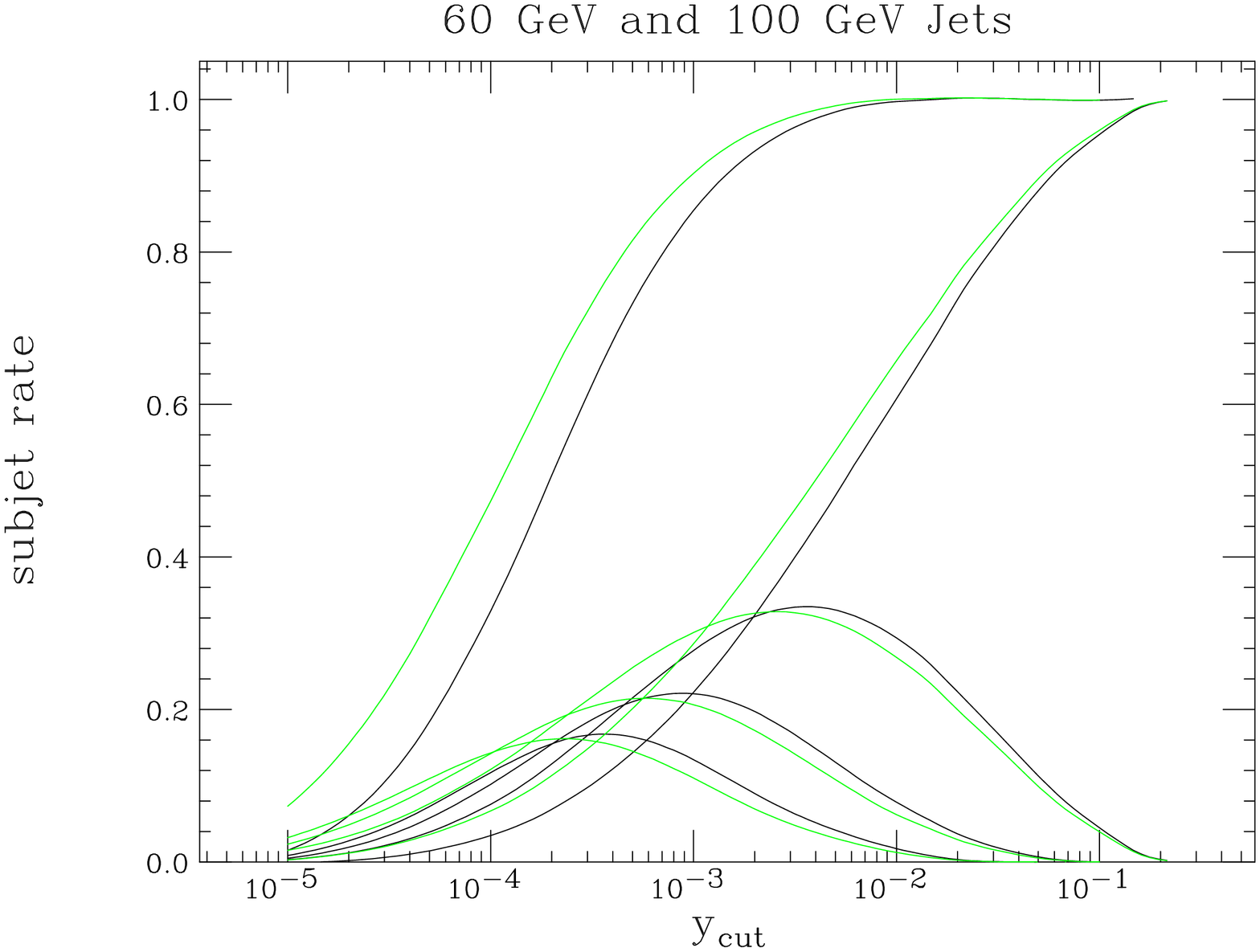}}}}
\caption{Comparison of subjet rates for jets of $p_t=60$ GeV (black)
and $p_t~=~100$~GeV (green).}
\label{ET}
\end{minipage}
\end{figure}

We turn now to the rapidity-dependence.  Figures~\ref{etaq}
and~\ref{etag} show the $\ycut$ dependence of central ($\eta=0$) and
forward ($\eta=2$) jets at $p_t = 60$ GeV and $\surd s = 1800$ GeV.
Again they are almost indistinguishable, while the rates for all
subjets, Figure~\ref{eta}, do vary owing to the differing mix of quark and
gluon jets.

\begin{figure}[ht]
\begin{minipage}[t]{0.475\textwidth}
\centerline{\resizebox{8cm}{!}{\rotatebox{90}{\includegraphics{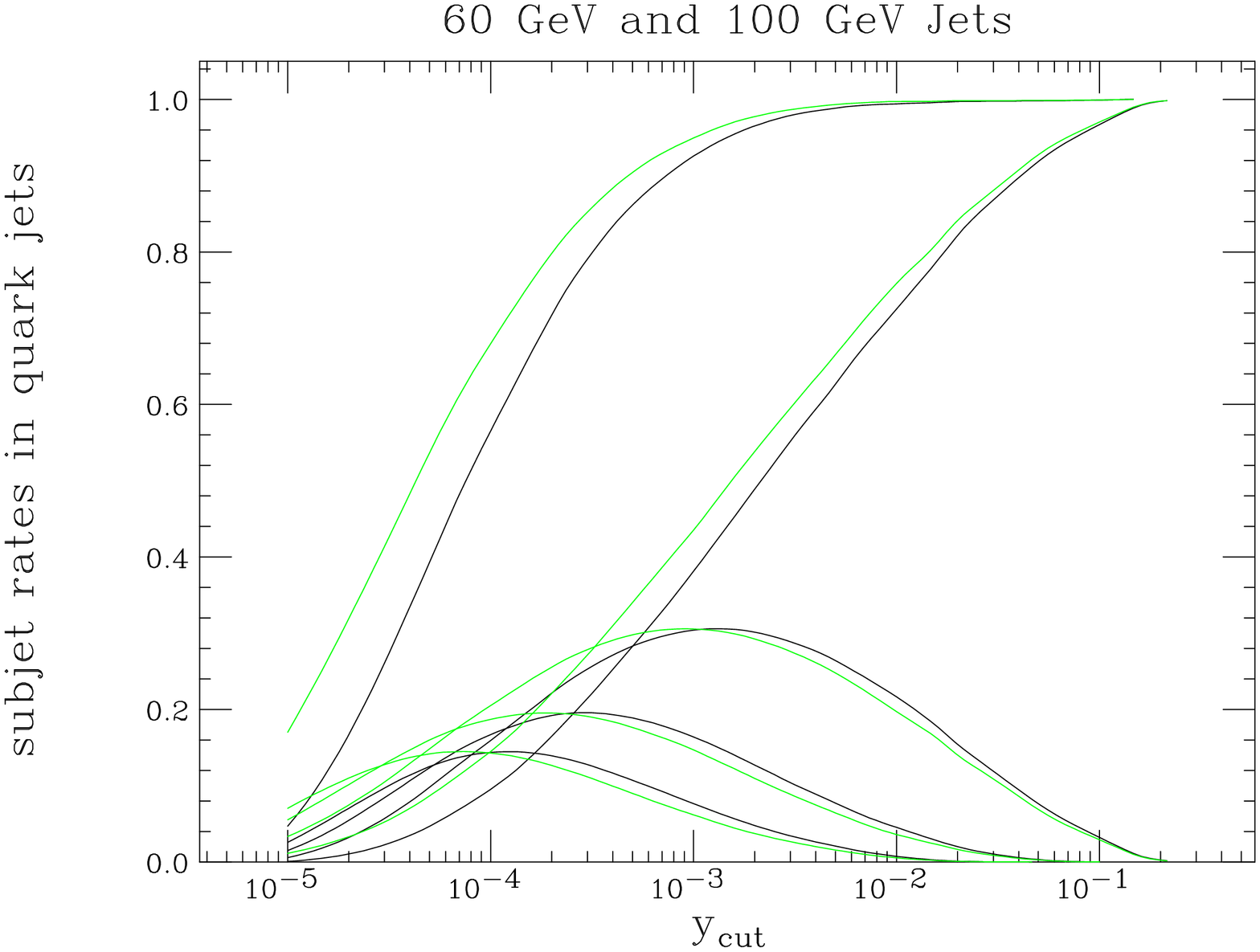}}}}
\caption{Comparison of subjet rates for quark jets of $p_t=60$ GeV (black)
and $p_t~=~100$ GeV (green).}
\label{ETq}
\end{minipage}\hspace*{\fill}
\begin{minipage}[t]{0.475\textwidth}
\centerline{\resizebox{8cm}{!}{\rotatebox{90}{\includegraphics{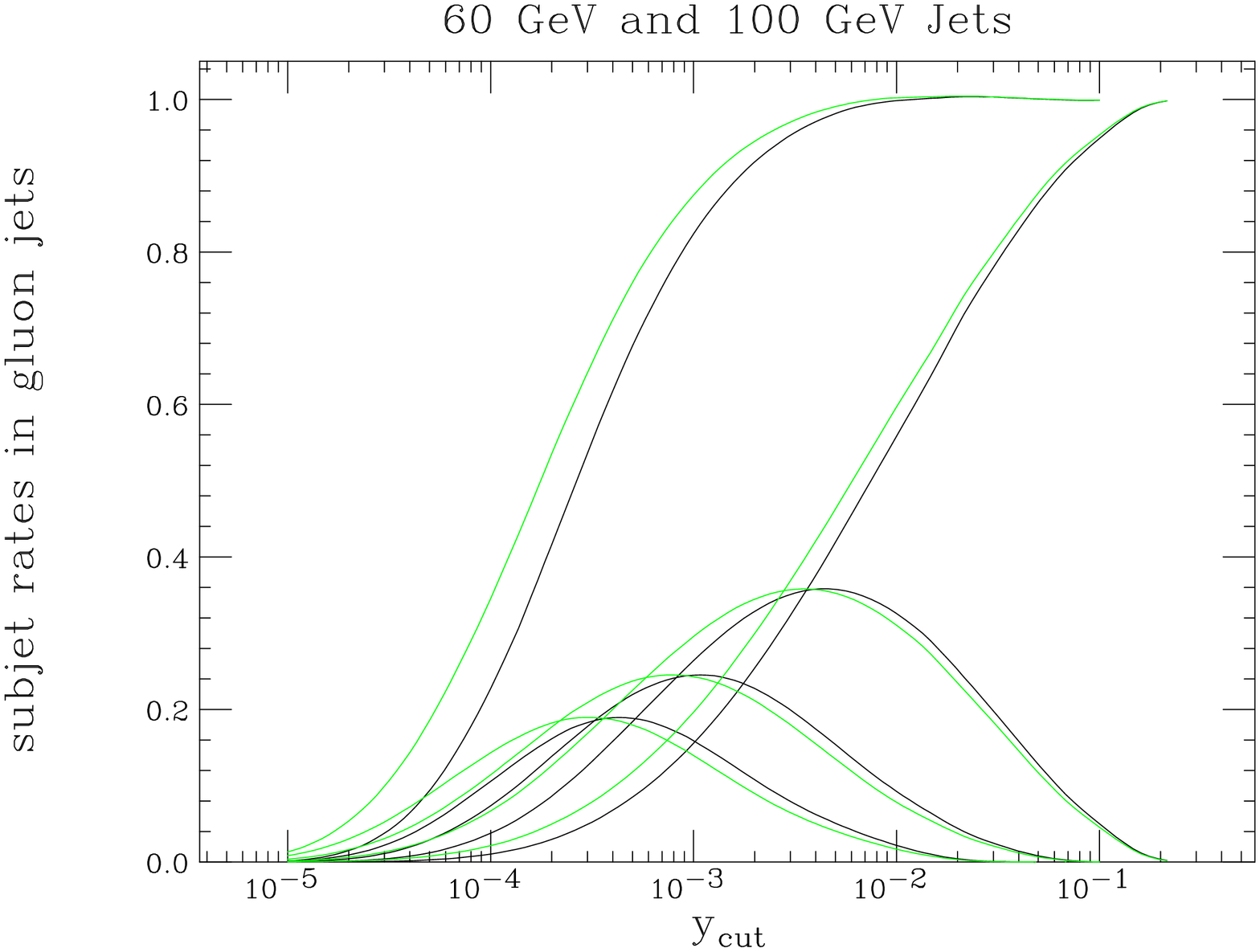}}}}
\caption{Comparison of subjet rates for gluon jets of $p_t=60$ GeV (black)
and $p_t~=~100$ GeV (green).}
\label{ETg}
\end{minipage}
\end{figure}

Figure~\ref{ET} shows the $\ycut$ dependence of $p_t=60$ GeV and
$p_t=100$ GeV jets at $\eta = 0$ and $\surd s = 1800$ GeV.  This time the
individual subjet rates for quark (Figure~\ref{ETq}) and gluon
(Figure~\ref{ETg}) jets do depend upon the jet $p_t$, because the initial
$\alps$ value is different.

\begin{figure}[ht]
\begin{minipage}[t]{0.475\textwidth}
\centerline{\resizebox{8cm}{!}{\rotatebox{90}{\includegraphics{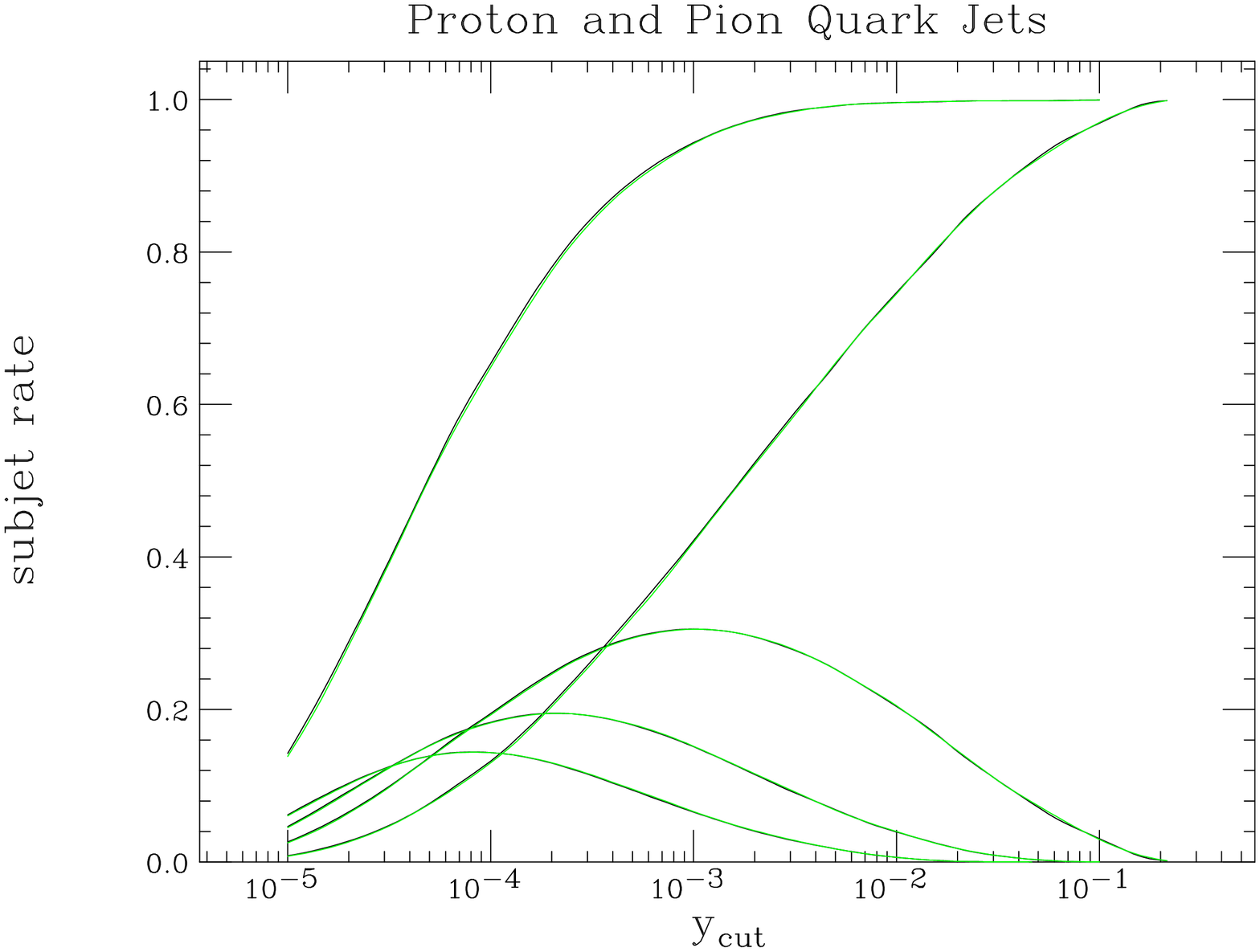}}}}
\caption{Comparison of subjet rates for quark jets in $p\bar{p}$ collisions
(black) and in $\pi^0 \pi^0$ collisions (green).}
\label{pionq}
\end{minipage}\hspace*{\fill}
\begin{minipage}[t]{0.475\textwidth}
\centerline{\resizebox{8cm}{!}{\rotatebox{90}{\includegraphics{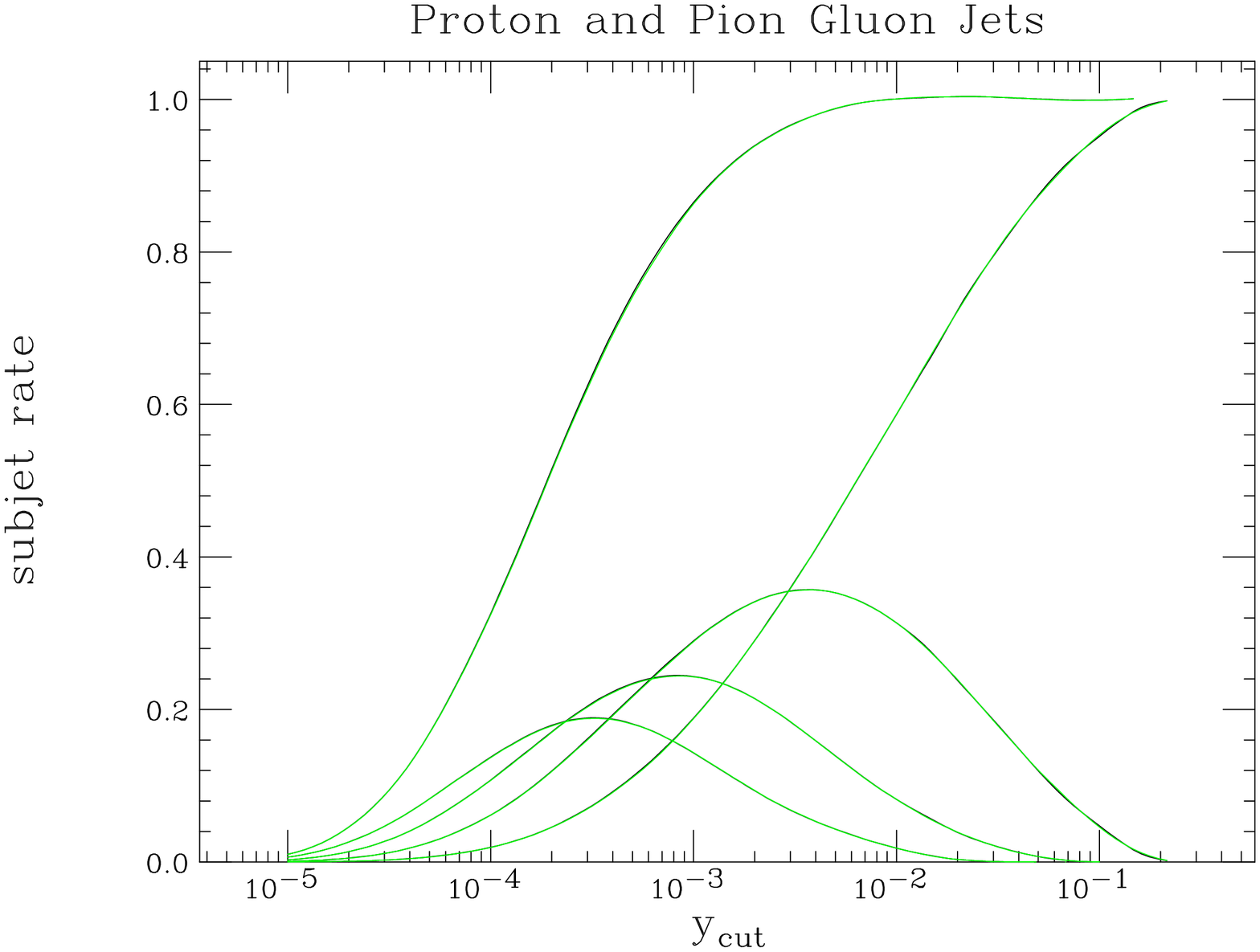}}}}
\caption{Comparison of subjet rates for gluon jets in $p\bar{p}$ collisions
(black) and in $\pi^0 \pi^0$ collisions (green).}
\label{piong}
\end{minipage}
\end{figure}
Finally, we show in Figures~\ref{pionq} and~\ref{piong} that not only are
the quark and gluon jet properties independent of the collision energy,
they are even independent of the collision type, by comparing canonical
$p \bar{p}$ jets with those from $\pi^0 \pi^0$ scattering
at the same energy.  To eliminate spurious differences due to different $\alps$
values in the different parton distribution functions,
we choose, for these plots only, the GRV sets~\cite{GRV,GRVP}, which use the
same $\Lambda$ value for both particle types.

This independence of the properties of a given flavoured jet from the
way it was produced is a highly non-trivial result.  One would normally
expect that the colour coherence of radiation from different emitters in
an event would make the soft radiation dependent on the full details of
the hard scattering.
As mentioned at the end of Section~\ref{fixed}, the amount of soft
initial-state radiation into the jet is in fact largely independent of
the scattering kinematics in the fully-inclusive case in which the
recoiling jet is integrated out.
However, putting requirements on additional jets breaks this
inclusivity and the colour coherence changes the properties of the
registered jet in response to the kinematics of the other jet.
\begin{figure}[ht]
\begin{minipage}[t]{0.475\textwidth}
\centerline{\resizebox{8cm}{!}{\rotatebox{90}{\includegraphics{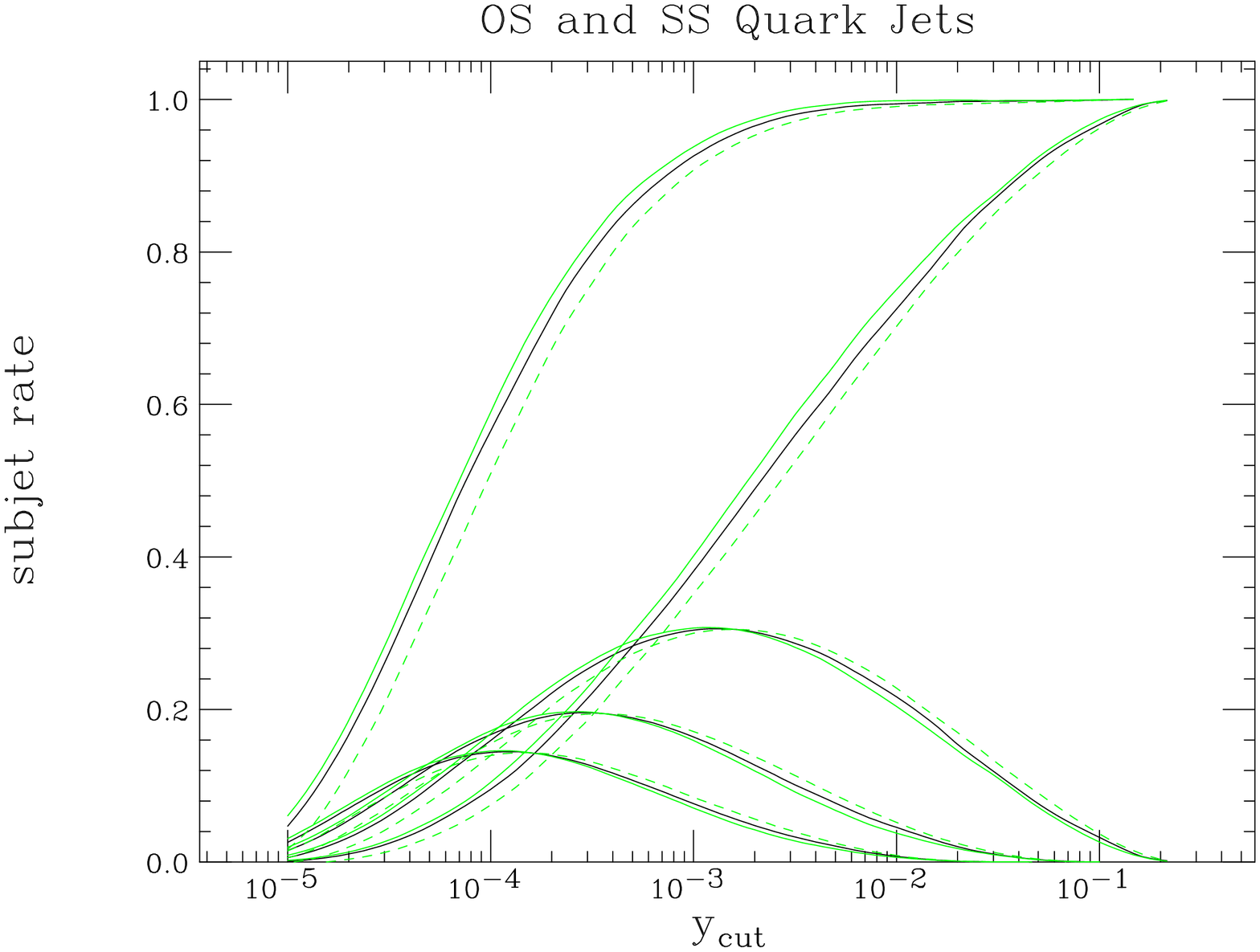}}}}
\caption{Comparison of subjet rates for quark jets at $\eta=2$ with the
other jet in the event allowed to be at any rapidity (black), required to
be at $|\eta|=2$ on the same side (solid green) or opposite side (dashed
green) of the event.}
\label{SSOSq}
\end{minipage}\hspace*{\fill}
\begin{minipage}[t]{0.475\textwidth}
\centerline{\resizebox{8cm}{!}{\rotatebox{90}{\includegraphics{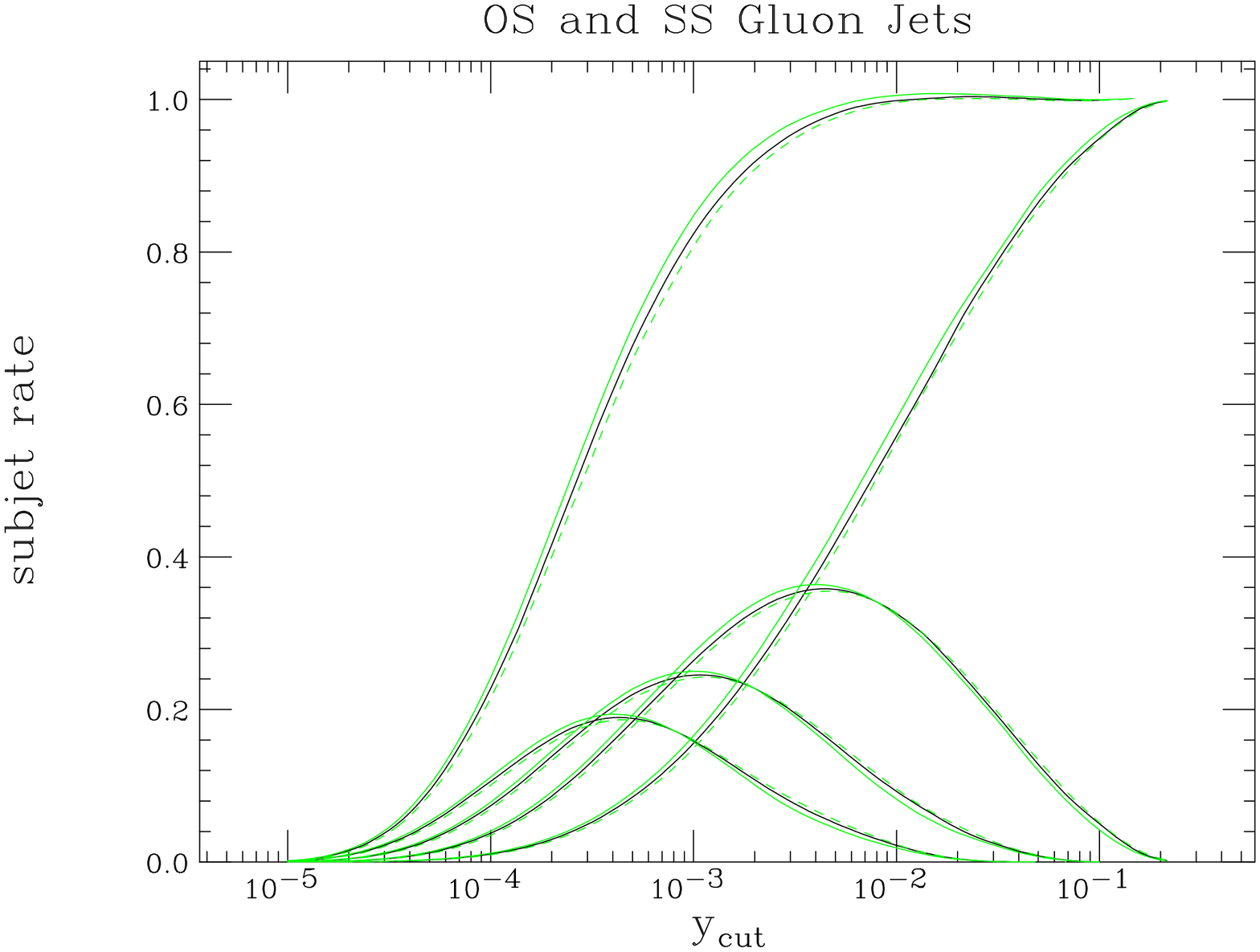}}}}
\caption{Comparison of subjet rates for gluon jets at $\eta=2$ with the
other jet in the event allowed to be at any rapidity (black), required to
be at $|\eta|=2$ on the same side (solid green) or opposite side (dashed
green) of the event.}
\label{SSOSg}
\end{minipage}
\end{figure}
This is demonstrated in Figures~\ref{SSOSq} and~\ref{SSOSg}, where we show
the quark and gluon subjet rates for jets at fixed rapidity in the
forward region, $\eta=+2$, with the recoiling jet either unconstrained,
or required to be at $\eta=+2$ `same side' or $\eta=-2$ `opposite side'.
The `same side/opposite side' ratio was once thought to be a good way to
separate quark and gluon jet properties since, for fixed jet kinematics,
the quark-to-gluon mix is very different in the two samples.
Figures~\ref{SSOSq} and~\ref{SSOSg} show that this is not the case, as
the selection itself considerably biases the jet properties.  Clearly
the method of \cite{DZ}, which uses the centre-of-mass energy-dependence
with fixed jet $E_T$ is superior.

Although we have not performed a calculation for photoproduction
($\gamma p$ collisions), which has a slightly different structure owing
to the direct photon contribution, we expect that the properties of
inclusively-defined jets there would be similar to those in hadron
collisions, which we have calculated.  However, because of the colour
coherence effect just mentioned, this statement is unlikely to be true
once cuts are made on the other jets in the event.  In particular, it is
common to try to separate experimentally the direct and resolved photon
events by imposing cuts on $x_\gamma$, the fraction of the photon's
momentum that is reconstructed in the hardest two jets in the
event.  It seems likely that this cut will bias the jet
properties sufficiently that our calculation cannot be used for a
quantitative analysis.

\section{Conclusion}
We have calculated the subjet rates in hadron collisions to
next-to-leading accuracy in logarithms of $\ycut$ to all orders in
$\alps$, matched with the exact $\cO(\alps)$ result at large $\ycut$.
To this accuracy the contribution from initial-state radiation is
essential.
Nevertheless, it is still possible to
separate quark and gluon jets, and we find that their properties are
almost completely independent of their production mechanism, depending
only on their~$p_t$, provided that they are defined fully inclusively.
As soon as additional cuts are placed on the event, the jet properties
become dependent on the details of the hard scattering.

Judging by comparisons of our results with and without threshold
matching, which formally differ only by uncalculated
next-to-next-to-leading logarithmic terms, we conclude that these terms
are rather large, especially for the higher subjet rates.  While this
situation could certainly be improved by matching the $n$-subjet rate to
the tree-level $n\!+\!1$-parton matrix element, further improvement will
be extremely difficult.  It is likely that, as in $e^+e^-$ annihilation,
$P_1$ will remain the best-calculated subjet rate.



\end{document}